\documentclass[10pt,conference]{IEEEtran}
\usepackage{graphicx} % Required for inserting images
\usepackage{xcolor}
\usepackage{soul}
\usepackage{braket}
\usepackage{amsmath}
\usepackage{amsfonts}
\usepackage{float}
\usepackage{subfigure} % For subfigures (alternative: use subcaption)

\title{On Selecting Paths for End-to-End Entanglement Creation in Quantum Networks}
\author{Anoosha Fayyaz,  Prashant Krishnamurthy, Kaushik P. Seshadreesan, David Tipper and Amy Babay\\ 
       Department of Informatics and Networked Systems \\
       School of Computing and Information \\
       University of Pittsburgh, Pittsburgh, PA 15260 USA \\
       {\it  email:} \{anf224, prashk, kausesh, dtipper, babay\}@pitt.edu }
\date{November 2024}

\begin{document}

\maketitle

\begin{abstract}
Optimal routing is a fundamental challenge in quantum networking, with several approaches proposed to identify the most efficient path for end-to-end (e2e) entanglement generation between pairs of nodes. In this paper, we show that \textit{prior entanglements} -- entanglements generated in a previous network cycle but not yet utilized -- are an important consideration in optimal path selection due to the dynamic nature of quantum networks. Specifically, we investigate whether a longer path with pre-existing entanglements can outperform a shorter path that starts from scratch. We account for key quantum constraints, including noisy entanglement generation and swapping, fidelity decay, probabilistic operations, and link discarding upon swap failure. Simulations reveal that longer paths with prior entanglements can establish e2e entanglement faster than shorter paths under certain conditions. We further introduce the notion of \textit{entanglement diversity}, where multiple paths can be used to improve performance -- either by selecting the first successful path to minimize time or using both paths to enhance fidelity through distillation. These findings highlight the importance of incorporating prior entanglements into path selection strategies for optimizing quantum communication networks.
    
\end{abstract}

\section{Introduction}
The rapid advancement of quantum computing has brought about a paradigm shift in how we approach computation and secure communication. To fully harness the power of quantum technologies, an underlying infrastructure is required to connect quantum devices across local and global scales. \textit{Quantum networks} provide this critical foundation by enabling the distribution of quantum information and entanglement across multiple nodes, facilitating applications such as quantum key distribution (QKD) \cite{bennett2014quantum,ekert1991quantum}, distributed quantum computing \cite{caleffi2018quantum}, and quantum-enhanced metrology \cite{yang2024quantum}. Unlike classical networks, where information is transmitted using bits encoded in electronic or optical signals, quantum networks rely on the transmission of quantum states encoded in photons.  

However, quantum networks face unique challenges that distinguish them from their classical counterparts. Quantum states are inherently fragile, and storing them in quantum memories leads to fidelity decay, reducing the quality of the qubits over time. Unlike classical packets, which can be duplicated and retransmitted, quantum states are governed by the no-cloning theorem \cite{wootters1982single}, which prohibits the replication of arbitrary quantum information. Once a qubit state has decohered, it is irretrievably lost, unlike classical packets that can be resent. Furthermore, the establishment of entangled links between quantum nodes is stochastic, with success probabilities dependent on factors such as the inter-nodal distance, noise, and hardware imperfections. Entangled links are also temporal in nature, decaying over time due to decoherence. Their quality is quantified by fidelity, which measures how closely the entangled state resembles a desired state.

Given these challenges, the most effective way to transfer quantum information between pairs of nodes is through quantum teleportation \cite{bennett1993teleporting}, a process that requires prior shared entanglement between communicating node pairs. This makes the \textbf{timely creation} of end-to-end (e2e) entanglement a critical task in quantum networks. Yet, the process of establishing e2e entanglement is stochastic, involving probabilistic entanglement generation over physical links, probabilistic swapping operations between entanglements that exist in adjacent links, and the management of decoherence in quantum memories. These factors make the design of efficient entanglement distribution policies a complex and challenging problem.

The existing literature has extensively explored optimization techniques for the distribution of entanglements, including Markov decision processes (MDPs) \cite{shchukin2019waiting,khatri2022design} and reinforcement learning approaches \cite{shchukin2022optimal}, with a focus on \textit{minimizing entanglement distribution time and maximizing entanglement generation rate}. However, many of these strategies assume idealized conditions, neglecting critical factors such as probabilistic entanglement generation, fidelity decay, memory cutoff times, and the impact of imperfect operations such as noisy entanglement generation and swapping. Moreover, most studies assume that the e2e entanglement generation process \textit{starts from scratch} for each entanglement distribution attempt, ignoring the dynamic nature of real-world quantum networks. When quantum networks are used repeatedly over time pre-existing (unused) entanglements from previous attempts at creating e2e entanglements remain. Such \textit{prior entanglements} can significantly influence the performance (rate or delay) of subsequent entanglement distribution tasks. Let us suppose that a ``path" is a set of physical links between a pair of nodes that need to transfer quantum information between each other. A path that is longest in terms of the number of such physical links might no longer be the ``longest" if some of its links have already established entanglements due to prior entanglement generation. This dynamic aspect of quantum networks is rarely addressed in the literature, as most studies assume a clean slate (entanglement-free links) for each e2e entanglement distribution attempt. %The problem of path selection in quantum networks is critical for the scalability and performance of quantum communication protocols. 
In a software-defined network where a controller with global view of the network can establish partial entanglements that can be swapped and stitched together to create e2e entanglements, knowledge of pre-existing entanglements is likely to improve the rate of entanglement generation. The decisions made by a network controller — such as where to establish entanglements and perform swaps --- depend on additional factors, including the physical distance, decoherence effects from quantum memories, the type of qubit architecture, and the fidelity threshold required for the application. 

In this paper, we take steps towards closing this gap of understanding what a network controller should do, by exploring the impact of pre-existing entanglements on the performance of e2e entanglement generation in quantum networks. Specifically, we compare two scenarios as a concrete case study: a clean-slate 2-hop path (no pre-existing entanglements) and a 4-hop path that benefits from pre-existing entanglements. We investigate under what conditions and for what values of entanglement generation probability ($p_e$) and swapping success probability ($p_s$) the 4-hop path outperforms the 2-hop path. By doing so, we aim to support a framework for entanglement distribution in quantum networks, taking into account the dynamic nature of network usage over time.

Our results demonstrate that for certain values of $p_e$ and $p_s$, the 4-hop path can establish e2e entanglement faster than the 2-hop path, particularly when pre-existing entanglements are considered. We refer to this as the \textit{4-hop favorable regime}. However, this advantage is highly dependent on the cutoff time, which determines how long entanglements can be stored before being discarded due to fidelity decay. We identify key thresholds for $p_e$ and $p_s$ where the 4-hop path outperforms the 2-hop path while ensuring e2e entanglement fidelity above the threshold. Further, we introduce the notion of \textit{e2e entanglement diversity} where a network controller can try more than one e2e entanglement establishment strategies towards improving the entanglement generation rate. The controller can adopt a \textit{first-completion policy} which improves this rate by selecting the first e2e entanglement with acceptable fidelity, agnostic of path. Alternatively, the controller can adopt an \textit{all-completion policy} where multiple e2e entanglements are created. As we explain in Section~\ref{sec:discussion}, if multiple e2e entanglements can be created, they can be \textit{distilled} to improve the overall fidelity of the e2e entanglement. Thus, even if one of multiple entanglements is slower or of lower fidelity, a quantum network may benefit from its establishment.

The remainder of this paper is organized as follows. Section II provides background on key concepts in quantum networking and summarizes related work. Section III presents the problem formulation and describes the network architecture and simulation setup. Section IV outlines the simulation results, while Section V analyzes these results in detail. Section VI discusses the implications of our findings and potential future directions. Finally, Section VII concludes the paper.

\begin{figure}[hth!]
    \centering
    \includegraphics[width=0.75\columnwidth]{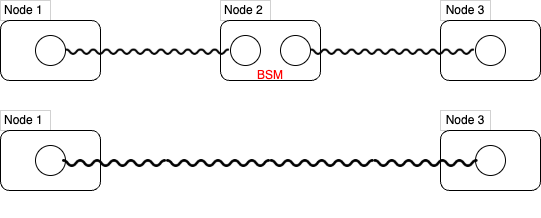}
    \caption{An entanglement swap performed between entangled pairs of nodes 1-2 and 2-3. A BSM is performed at the intermediate node (node 2), resulting in the connection of nodes 1 and 3 without requiring direct interaction between them}
    \label{fig:swap}
\end{figure}

\section{Background}

\subsection{Entanglement, Swapping, and Teleportation}
At the core of quantum networking is quantum entanglement, a non-classical correlation between quantum particles that persists even when the particles are separated by large distances. In quantum networks, entanglement serves as the fundamental resource for enabling communication protocols. For instance, quantum teleportation (discussed below) relies on pre-shared entanglement to transmit quantum states without physically transferring the qubits \cite{bennett1993teleporting}. Similarly, QKD protocols like BB84 and E91 exploit entanglement to achieve unconditional security. Efficient entanglement distribution and maintenance is essential for scalable quantum communication, yet it presents significant physical and engineering challenges.

%Entanglement is a fundamental quantum phenomenon wherein two or more qubits exist in a shared quantum state, such that their individual states cannot be described independently of each other. Unlike classical bits, which exist deterministically in states $0$ or $1$, qubits can exist in superpositions and exhibit non-classical correlations that defy local realism. 
Mathematically, an entangled state of two qubits, A and B, can be expressed as
\[
    \ket{\psi}_{AB} = \alpha \ket{0}_A \ket{0}_B + \beta \ket{1}_A \ket{1}_B = \alpha \ket{00}+\beta\ket{11},
\]
where $|\alpha|^2 + |\beta|^2 = 1$. These qubits cannot be written as a tensor product unlike product states, such as $\ket{\psi} = \ket{0} \otimes \ket{1}$ that describe independent qubits with no quantum correlations. 
%The key distinction lies in the measu therement outcomes, where measuring one qubit instantaneously determines the state of the other for an entangled state, regardless of their spatial separation. This non-local property enables applications in secure communication, quantum computing, and quantum-enhanced sensing.

Entanglement swapping extends the reach of quantum correlations by creating entanglement between two qubits that have never interacted directly. Suppose two independent entangled pairs $AB$ and $CD$ are prepared in states
\[
    \ket{\psi}_{AB} = (\alpha \ket{00}+\beta\ket{11})_{AB}, \ket{\phi}_{CD} = (\gamma \ket{00}+\delta\ket{11})_{CD}.
\]
If an intermediate joint Bell-state measurement (BSM) is performed on qubits B and C, their states collapse, and qubits A and D become entangled, effectively establishing long-range entanglement as shown in Fig. \ref{fig:swap}. However, practical implementations are noisy, leading to imperfect fidelity. The fidelity $\mathcal{F}$ of the resulting entangled state with respect to the ideal state is given by:
\begin{equation}
    \mathcal{F} = \bra{\psi_{AD}} \rho_{AD} \ket{\psi_{AD}},
\end{equation}
where $\rho_{AD}$ is the density matrix of the final state and $\ket{\psi_{AD}}$ is the desired state. Higher fidelity indicates closer resemblance to the ideal state, but decoherence and operational imperfections degrade this value. Entanglement swapping is a key primitive for enabling long-distance quantum communication and distributed quantum computing \cite{buhrman2003distributed, caleffi2024distributed}.

Quantum teleportation is a protocol that allows the transmission of an arbitrary quantum state using entanglement and classical communication. Suppose Alice wants to teleport an unknown qubit state $\ket{\psi}$ and shares an entangled state $\ket{\phi^+_{AB}}$ with Bob. The combined initial state of the system is $\ket{\psi} \otimes \ket{\phi^+{AB}}$. Alice performs a Bell-state measurement (BSM) on the unknown qubit and her entangled qubit A, projecting them into one of the four Bell states. Bob’s qubit is correspondingly projected into one of four possible states, each related to $\ket{\psi}$ by a known unitary transformation. Alice then communicates the outcome of her measurement to Bob using two classical bits.
Upon receiving Alice’s two-bit message, Bob applies an appropriate unitary operation $U_B$ to recover the original state $\ket{\psi}$. Similar to entanglement swapping, the fidelity of teleportation depends on the quality of the initial entangled pair. Quantum teleportation enables secure quantum communication and plays a key role in quantum networking by allowing state transfer without direct transmission, mitigating decoherence effects in long-distance communication channels.

%Unlike classical bits, which exist deterministically in states $0$ or $1$, qubits can exist in superpositions of these states, described as $\ket{\psi} = \alpha \ket{0} + \beta \ket{1}$. However, the inherent fragility of quantum systems poses significant challenges for quantum networking. Qubits are susceptible to decoherence, a process in which interactions with the environment degrade their quantum properties over time. Consequently, maintaining high-fidelity entanglement across a network requires overcoming the dual challenges of noise and probabilistic success in entanglement generation and operations.

\subsection{Quality of Entanglements}
One of the fundamental obstacles in quantum networks is the rapid loss of quantum coherence over long distances due to environmental interactions, including decoherence from noise (e.g., thermal, electromagnetic, and spin noise), coupling to surrounding systems, and operational imperfections in quantum hardware. Unlike classical signals, quantum states cannot be amplified due to the no-cloning theorem, which prohibits the deterministic duplication of arbitrary quantum information \cite{wootters1982single}. As a result, direct transmission of entangled photon pairs across long distances becomes exponentially inefficient. 
To circumvent this limitation, quantum repeaters have been proposed to extend the range of entanglement distribution. These devices operate by breaking the long-distance communication into shorter segments, generating entanglement between neighboring nodes, and performing entanglement swapping to connect these segments \cite{briegel1998quantum, azuma2023quantum}.

While quantum repeaters mitigate the exponential loss of photons in optical fibers, they introduce new challenges. Entanglement generation and swapping are probabilistic operations, with success rates influenced by factors such as channel noise, memory decoherence, and imperfect measurements. Moreover, the storage of entangled states in quantum memories is time-sensitive, as decoherence degrades the fidelity of the stored states over time, necessitating error correction or entanglement distillation protocols. Furthermore, existing repeater protocols typically rely on simplistic entanglement distribution policies, such as swapping entanglements as soon as two links are available, without accounting for global network conditions. These challenges create a trade-off between the speed of entanglement distribution and the quality of the distributed entanglement, making the design of efficient entanglement distribution policies a critical research problem.

\subsection{Related Work}

Optimal routing and policies for entanglement distribution in quantum networks have been studied in recent years, with a focus on maximizing efficiency, minimizing latency, and ensuring robust performance under varying network conditions. Markov Decision Processes (MDPs) have been extensively utilized to model waiting times for end-to-end entanglement generation and to develop optimal policies for entanglement distribution \cite{shchukin2019waiting, khatri2022design, shchukin2022optimal}. Recent advancements have further integrated memory cut-off times into these models, enabling optimized policies for homogeneous repeater chains \cite{inesta2023optimal}.
%though pre-existing entanglements were not considered. 
Fidelity optimization has been focused as well, with studies establishing fidelity thresholds for elementary links \cite{khatri2021policies}, determining the optimal sequence of distillation and swapping operations in two-link linear networks \cite{koutsopoulos2024optimal}, and designing distillation-based routing strategies for multiple source-destination pairs \cite{li2022fidelity}. Additionally, multi-path strategies to enhance distribution rates for multiple users have been explored \cite{sutcliffe2023multiuser}. Research on optimal routing design has utilized metrics such as the average end-to-end entanglement rate \cite{caleffi2017optimal}, while evaluating whether incorporating a repeater or relying on a direct link would yield better performance. Quantum overlay networks have been proposed to optimize entanglement distribution across multiple demands by storing entanglements during low-demand periods for use during high-demand times \cite{pouryousef2023quantum}. Furthermore, swapping trees have been investigated to identify optimal entanglement swapping locations, thereby reducing end-to-end entanglement generation latency \cite{ghaderibaneh2022efficient}. \textit{Despite these advancements, existing studies often rely on idealized assumptions, overlooking at least one of the critical factors such as probabilistic entanglement generation and swapping, noisy entanglement generation and entanglement swapping operations, and memory cut-off times due to fidelity decay}. Moreover, most of these works do not account for pre-existing entanglements in the network, which can be  beneficial for practical quantum networks. In dynamic network environments, the network does not always start from scratch, meaning longer paths may not necessarily be less efficient when prior established entanglements are considered. To the best of our knowledge, our work is the first to integrate all of these factors into a comprehensive framework for e2e entanglement generation. Notably, we explicitly include prior entanglements in the analysis in addition to all the physical constraints.

\begin{figure}[hth!]
    \centering
    \includegraphics[width=0.85\columnwidth]{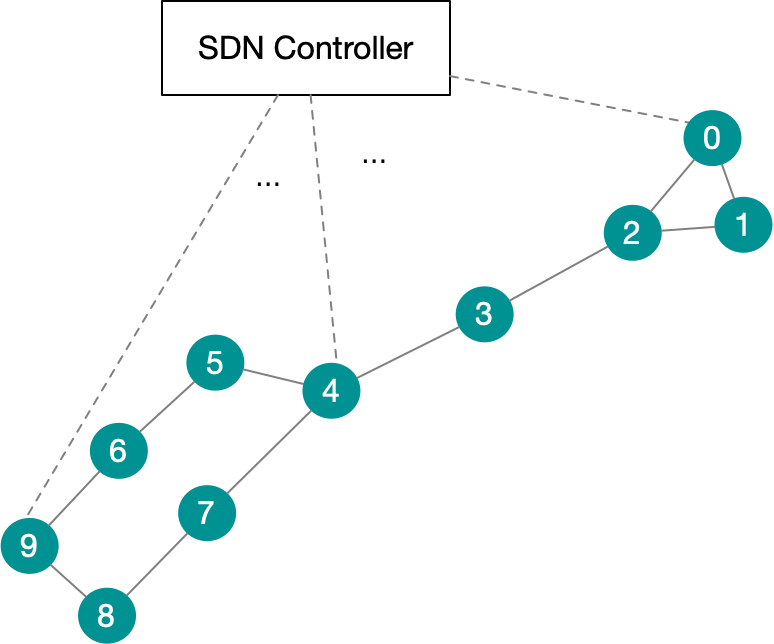}
    \caption{Network of $n=10$ nodes and an SDN controller}
    \label{fig:network}
\end{figure}

\section{Formulating the case study: 2-hop path vs 4-hop path}

\subsection{Network Architecture}

\begin{figure}
    \centering
    \includegraphics[width=0.95\linewidth]{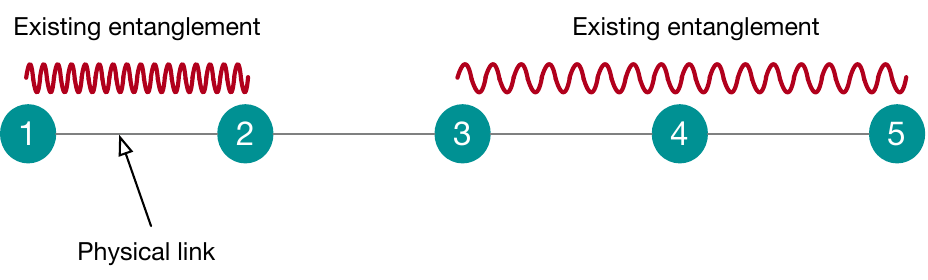}
    \caption{Path with four hops with partial entanglements between nodes 1-2 and 3-5}
    \label{fig:fourhops}
\end{figure}

We consider a network consisting of $n$ nodes, where certain nodes aim to establish entanglements for the purposes of information sharing or communication (see Fig.~\ref{fig:network}). A Software-Defined Networking (SDN) controller possesses a comprehensive view of the network at any given time $t$. The SDN controller directs the nodes on the creation of entanglements, swapping operations, and the creation of end-to-end entanglement to enable subsequent exchange of quantum information through teleportation. We focus here on two distinct paths between the same end nodes. One path consists of two hops with no prior shared entangled pair, while the other has four hops and includes some pre-established entanglements from a prior cycle. In the four-hop path, entanglements are already established between nodes 1-2 and 3-5 as shown in Fig. \ref{fig:fourhops}.

%Classical connections between all nodes are assumed to be pre-established, while quantum links will be created in the form of quantum entanglements. In a linear chain, each intermediate node establishes entanglement with both of its neighboring nodes simultaneously, limiting the memory requirement to just two qubits. Additional memory for intermediate nodes may be considered when the case study is extended to accommodate varying network demands. However, the end nodes require only memory for a single qubit. Attempting end-to-end entanglement through two distinct paths for the same source-destination pair may necessitate two-qubit memory at the end nodes, as is the case for \textit{entanglement diversity} that we explain later. 

We assume that all nodes are equipped with identical hardware for entanglement generation, meaning the same physical setup (such as an SPDC-based source) is used uniformly across the network. In such setups, the entanglement source is located between two nodes, where entangled photons are generated and sent to the nodes, with each node storing its respective photon in a quantum memory upon successful detection, thus establishing a shared bipartite entanglement between them. Since the probability of successful link generation depends on photon loss over distance, assuming a constant link generation probability $p_e$ implies that all inter-nodal distances are equal. Similarly, we assume identical hardware for entanglement swapping, leading to a uniform swapping success probability $p_s$ across the network. These assumptions ensure that the 4-hop path is longer than the 2-hop path in terms of the total distance covered\footnote{Thus, the 4-hop path is inherently at a disadvantage. What we see is that under certain conditions, it still generates e2e entanglement faster than a 2-hop path when partial entanglements are considered.}. Alternatively, one could consider varying link lengths, which would result in different probabilities of entanglement generation for each link. However, such a scenario is not necessary for our purpose, which is to demonstrate that, under certain conditions, a longer path with pre-existing entanglements may establish e2e entanglement faster than a shorter path starting from scratch. To maintain generality and accommodate different quantum technologies and entanglement generation sources, we adopt the assumption of uniform entanglement generation probabilities across all links.

Classical connections between all nodes are assumed to be pre-established, while quantum links will be created in the form of quantum entanglements. In a linear chain, each intermediate node establishes entanglement with both of its neighboring nodes simultaneously, limiting the memory requirement to just two qubits. Additional memory for intermediate nodes may be considered when the case study is extended to accommodate varying network demands. However, the end nodes require only memory for a single qubit. Attempting end-to-end entanglement through two distinct paths for the same source-destination pair may necessitate two-qubit memory at the end nodes, as is the case for entanglement diversity that we explain later. 
Additionally, we assume that all nodes are equipped with identical quantum memories, which follow the same noise model. 

\subsection{Noise Model and Fidelity}
The quantum channels connecting the nodes are modeled as depolarizing channels, a common approach in the literature to account for noise and decoherence in quantum communication. The creation of bipartite entanglements and entanglement swapping operations are considered to be noisy. Since the newly created state is not a maximally entangled state, we consider them to be Werner states, a noisy state which is a mixture of a maximally entangled Bell state and a maximally mixed state. Werner states are represented as 
\begin{equation}
    \label{Werner}
    \rho = \frac{4\mathcal{F}-1}{3} \ket{\phi^+}\bra{\phi^+} + \frac{1-\mathcal{F}}{3} \mathbb{I}_4
\end{equation}
where $\mathcal{F}$ is the fidelity with which the Werner state resembles the Bell state $\ket{\phi^+}$.

The noise in quantum memories can be modeled as a depolarizing channel of the form \cite{nielsen2010quantum}
\begin{equation}
    \label{depo_channel}
    \mathcal{E}(\rho) = p\frac{I}{2} + (1-p) \rho,
\end{equation}
where $\rho$ is the quantum state on which the depolarizing channel acts, and $p$ is the probability with which the state becomes maximally mixed. 

The fidelity drop in quantum memories is given by \cite{inesta2023optimal}
\begin{equation}
    \mathcal{F}(t) = \frac{1}{4}+ \left( \mathcal{F}_0 - \frac{1}{4} \right)  e^{-t/\tau}
\end{equation}
where $\mathcal{F}_0$ is the inital fidelity, $\tau$ is the decoherence time constant of the memory and $t$ is the elapsed time. To model the imperfect entanglement creation operation, we assume that the newly created entanglement has unit fidelity but is subjected to one application of the depolarizing channel so the above equation can be adjusted to become
\begin{equation}
    \label{F(t}
    \mathcal{F}(t) = \frac{1}{4}+ \left( \mathcal{F}_0 - \frac{1}{4} \right)  e^{-(t+1)/\tau}
\end{equation}
and $\mathcal{F}_0$ can be effectively considered $1$.
The fidelity of the resulting entanglement after a successful swap is given by \cite{sen2005entanglement}
\begin{equation}
    \label{F_swap}
    \mathcal{F}_{12} = \frac{1}{4} + \frac{3}{4} \left( \frac{4\mathcal{F}_1-1}{3} \right)\left( \frac{4\mathcal{F}_2-1}{3} \right),
\end{equation}
where $\mathcal{F}_1$ is the fidelity of the first entangled pair and $\mathcal{F}_2$ is the fidelity of the second entangled pair.

% Atomic ensemble memories: Milliseconds to seconds (T_d ~ 1ms to 1000ms)
%Cavity-based systems: Microseconds to millisecond (T_d ~ 1ms - 100ms)

\subsection{Network Policies}

\begin{figure}
    \centering
    \includegraphics[width=0.95\linewidth]{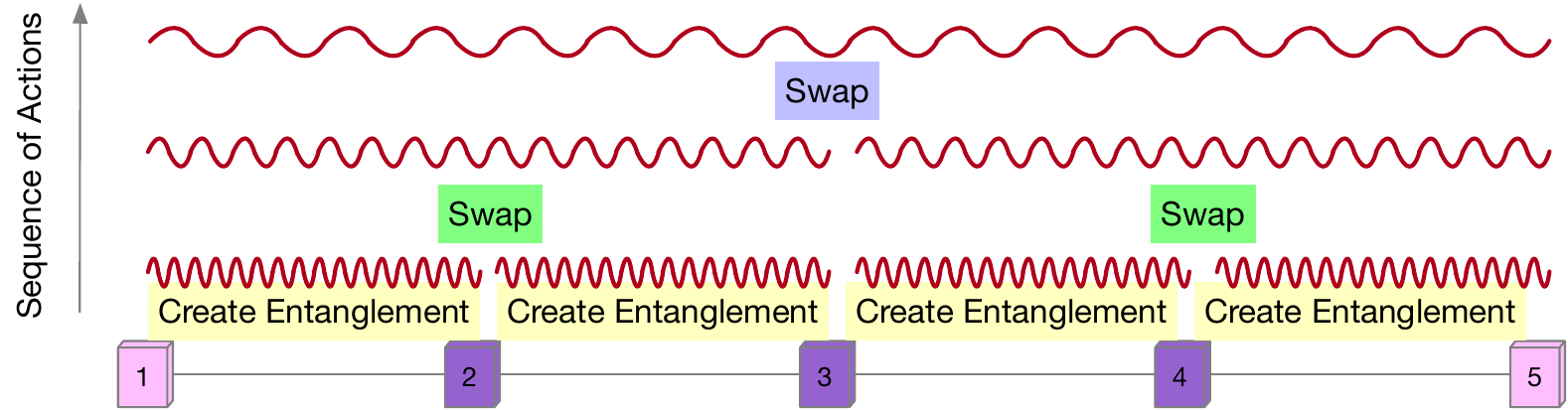}
    \caption{Sequential swaps}
    \label{fig:swaps}
\end{figure}

Time is discretized into time steps ($\Delta t$), each sufficient for one entanglement generation or swap operation. The duration of a time step is determined by the maximum time required for either an entanglement trial or a swap operation, including any associated overhead, such as cooling or reinitialization. Mathematically, this is expressed as
\begin{equation}
    \Delta t = \text{max}(t_{\textrm{e, misc}}, t_{\textrm{s, misc}}),    
\end{equation}
where $t_{e, \text{misc}}$ and $t_{s, \text{misc}}$ represent the total time required for entanglement generation and swap operations, respectively, including 
additional preparation time in case of a failed attempt. Nodes with available memory attempt entanglement creation in each time step. We do not assume instantaneous classical communication so swaps for entanglements generated within a time step are deferred to the next time step. For linear chains, swaps are performed in pairs. If entanglements are generated in adjacent pairs in a previous time step, the nodes would attempt a swap. For instance, in a 4-hop chain with entanglements between nodes 1-2, 2-3, 3-4, and 4-5, swaps occur first at nodes 2 and 4, followed by node 3, as illustrated in Fig.~\ref{fig:swaps}. If only nodes 1-2 and 3-4 share entanglements, swaps are delayed until adjacent pairs complete their entanglements. For more complex networks, swapping trees can be considered \cite{ghaderibaneh2022efficient}. Waiting to swap for this case results in a lesser entanglement fidelity than the swap-asap policy and then wait for other entanglements to be created. The fidelity associated with the waiting for a time $t_w$ to swap is given by 
\begin{equation}
    \label{F_wait_swap} 
    \mathcal{F}_{\textrm{wait-swap}} = \frac{3}{4} \left[ \frac{1}{3} + \left(\frac{4\mathcal{F}_{01}-1}{3}\right)\left(\frac{4\mathcal{F}_{02}-1}{3}\right) e^{-2 t_w/\tau} \right],
\end{equation}
and the swap-asap and then wait for $t_w$ time steps is given by 
\begin{equation}
    \label{F_swap_wait}
    \mathcal{F}_{\textrm{swap-wait}} = \frac{3}{4} \left[ \frac{1}{3} + \left(\frac{4\mathcal{F}_{01}-1}{3}\right)\left(\frac{4\mathcal{F}_{02}-1}{3}\right) e^{-t_w/\tau} \right],
\end{equation}
where $\mathcal{F}_{01}$ is the initial fidelity of the first entanglement and $\mathcal{F}_{02}$ is the initial fidelity of the second entanglement, similar to \cite{inesta2023optimal}. Since $\mathcal{F}_{wait-swap} < \mathcal{F}_{swap-wait}$, the swap-asap policy yields higher fidelity. By assuming the worst-case scenario (waiting to swap), we do not lose generality, and this assumption does not affect the conclusion that the 4-hop path can outperform the 2-hop path under certain conditions. 

To ensure that the fidelity of the e2e entanglement between the end nodes remains above a specified threshold, we introduce \textbf{a cutoff time}, $T$. Entanglements older than this cutoff time are discarded to prevent the e2e fidelity from falling below the desired threshold. Based on Eq.~\ref{F(t}, the cutoff time $T$ is defined as
\begin{equation}
    \label{F_th}
    T = -\tau \ln\left(\frac{4\mathcal{F}_{th}-1}{4\mathcal{F}_0-1} \right) -1,
\end{equation}
where $\mathcal{F}_0 =1$ and $\mathcal{F}_{th}$ is the fidelity threshold. All of the quantities of time are measured in discrete time steps. %Typically $\tau$ is orders of magnitude larger than $T$. 

\begin{figure*}[h!t]
    \centering
    {\includegraphics[width=0.24\textwidth]{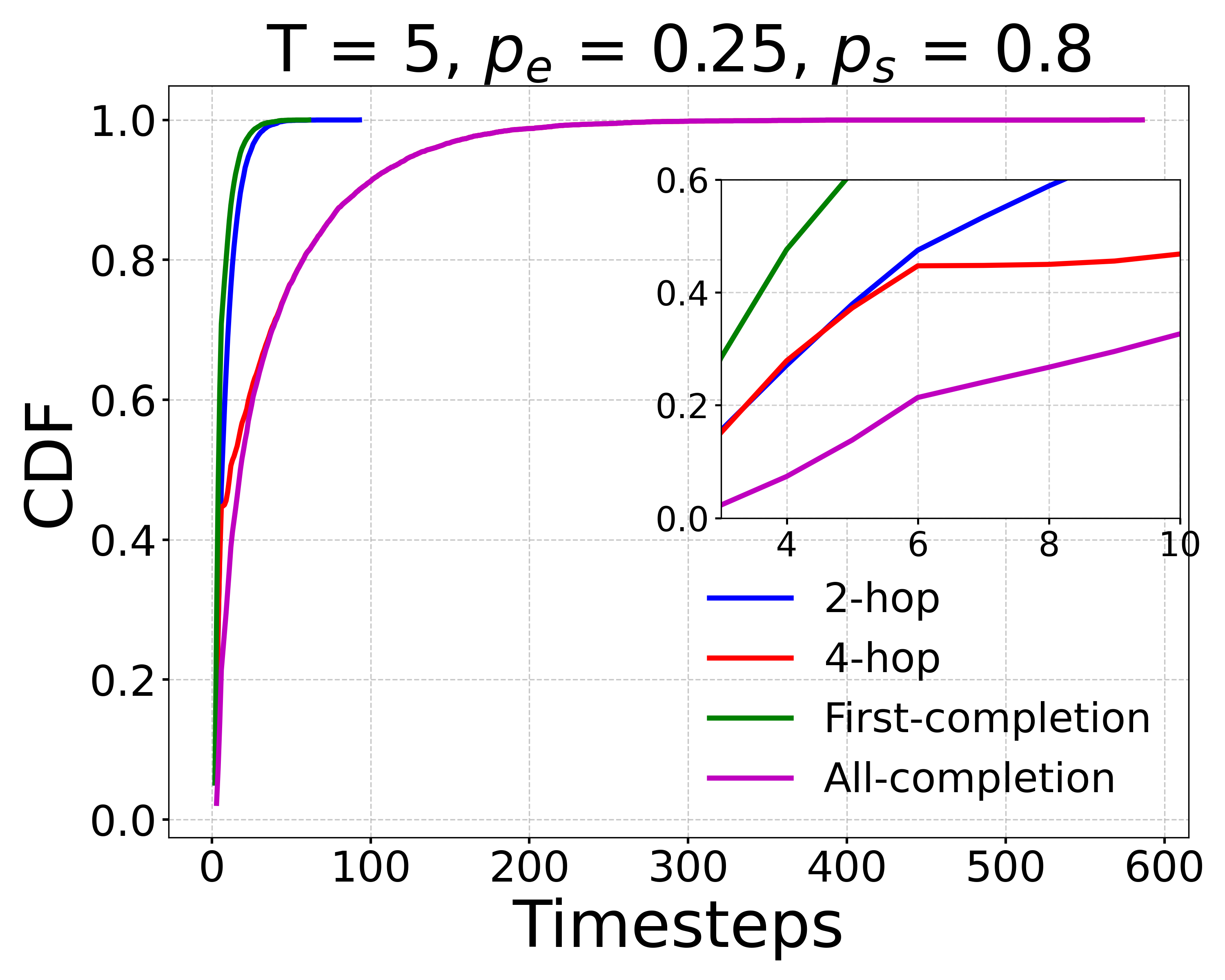}} 
    {\includegraphics[width=0.24\textwidth]{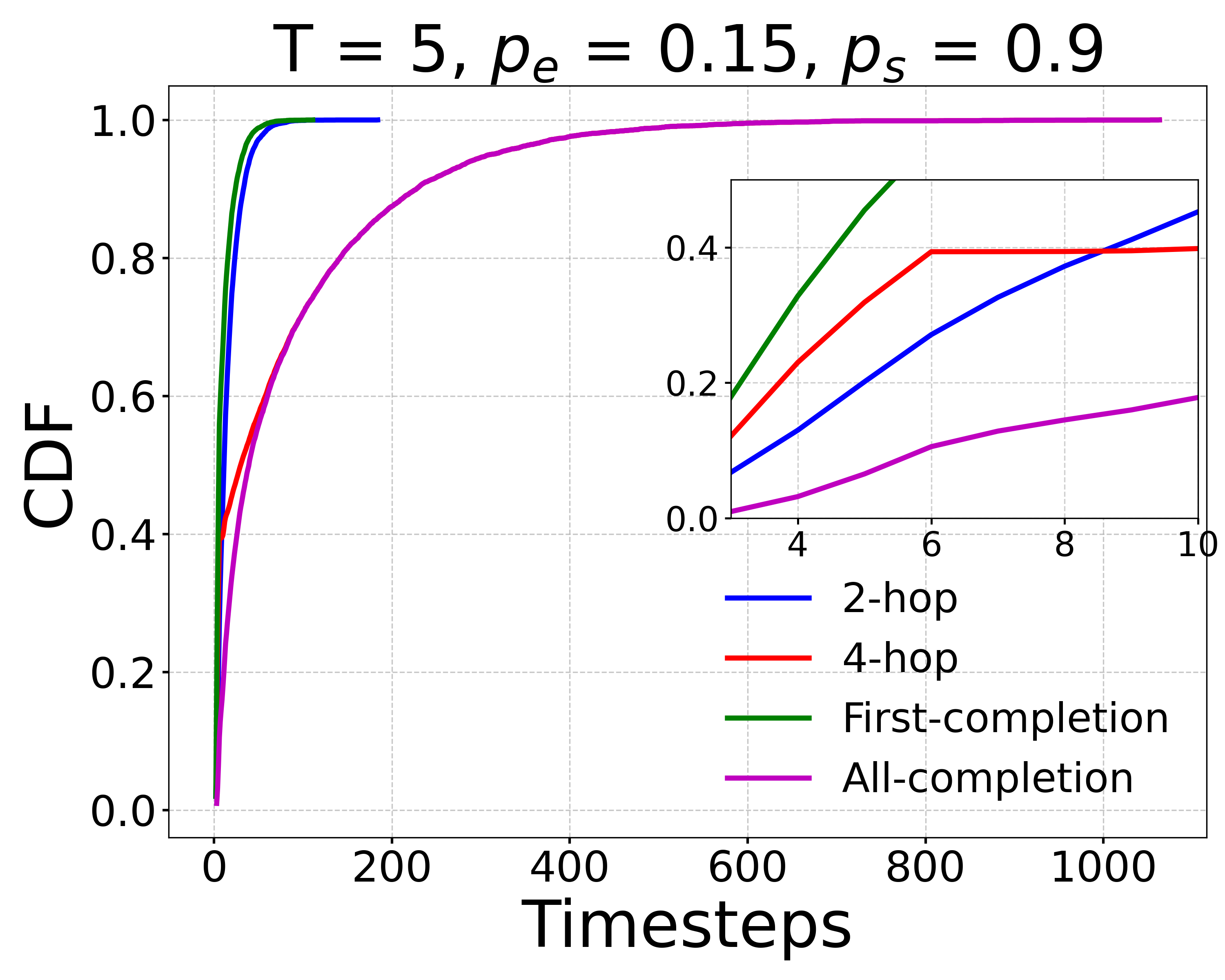}} 
    {\includegraphics[width=0.24\textwidth]{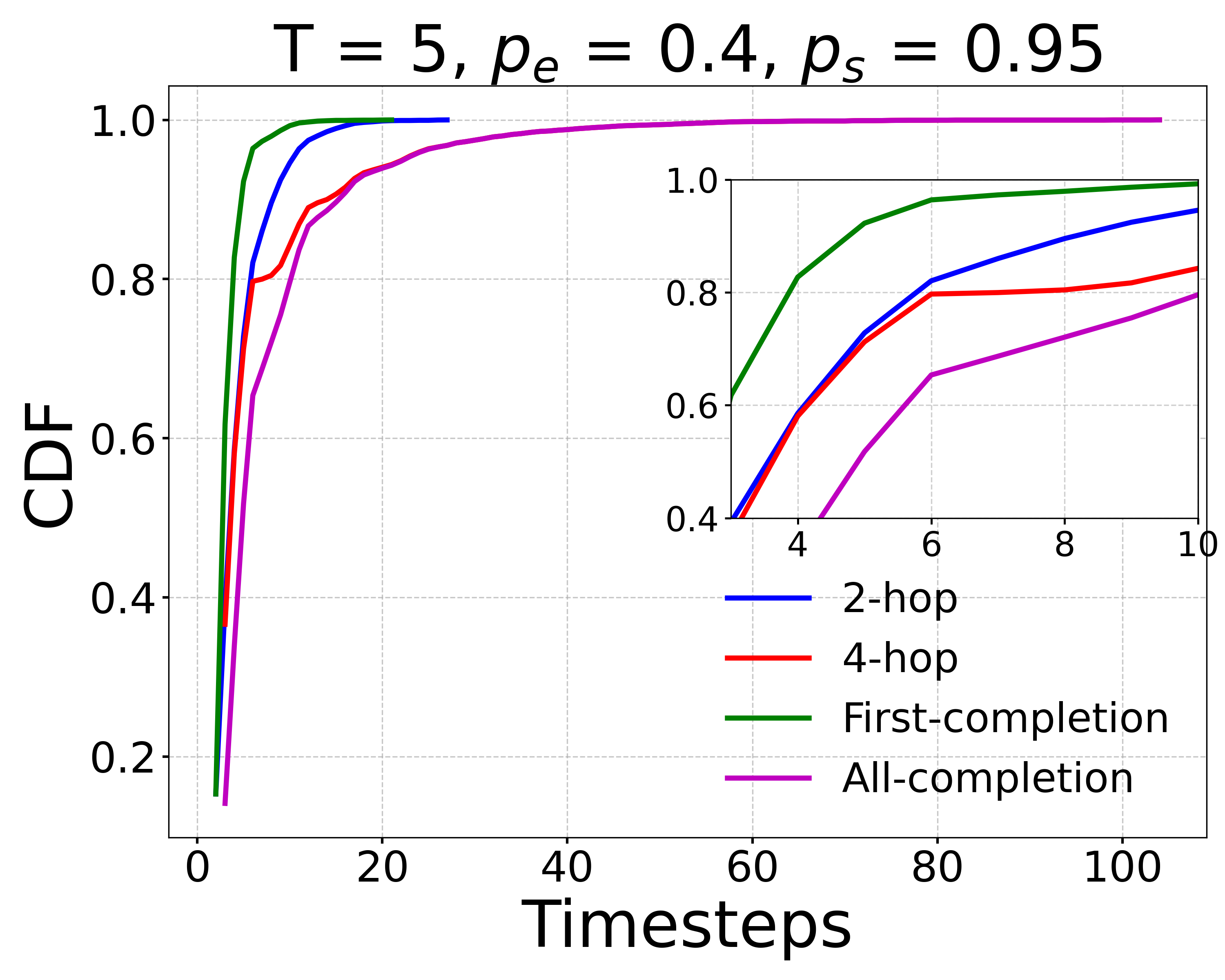}} 
    {\includegraphics[width=0.24\textwidth]{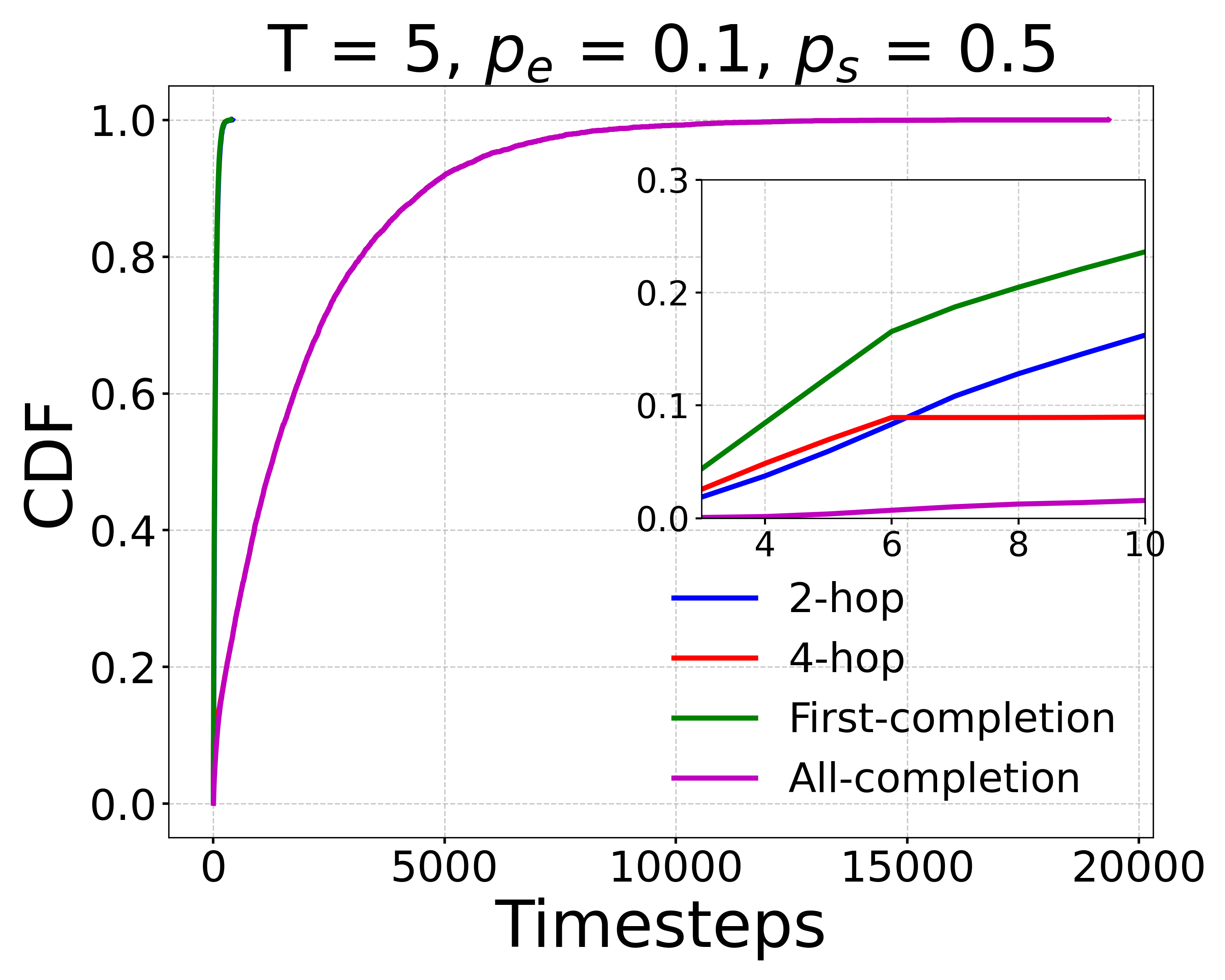}} 
    \caption{CDF of the number of time steps required for e2e entanglement generation for various values of $p_e$ and $p_s$ with $T$ = 5}
    \label{T=5 figures}
\end{figure*}

\subsection{Entanglement Diversity}

We examine two policies in our work that support a new notion we call \textit{Entanglement Diversity} for e2e entanglement generation in quantum networks. Diversity has been extensively used in wireless communications where signal fades can increase bit errors (see for e.g., \cite{Tse04, Neas98}). In the simplest case, many receiver antennas get copies of signals with different (uncorrelated) fading characteristics. In selection diversity, the receiver selects the best received signal among the multiple fading copies, discarding the others\footnote{The probability that all $k$ copies have poor quality ($p^k$) is lower than the probability $p$ that a single received signal has poor quality.}. Analogous to selection diversity, we introduce the ``\textit{first-completion}" policy. In this case, two alternative (disjoint -- in this paper) paths are attempted to generate e2e entanglement and the time is recorded as soon as one of the paths manages to establish e2e entanglement. The SDN controller discards the entanglement generation process in the alternative path. In a network (beyond this isolated example), this approach is advantageous when there are no other demands competing for the utilization of edges along the two paths involved. In such cases, it enables significantly faster e2e entanglement generation (since the process is probabilistic). Moreover, even in the presence of other demands, the entanglements created in the path that does not achieve e2e entanglement first can be repurposed as prior entanglements for other paths. This creates a win-win situation, as it not only accelerates entanglement generation for the current task but also enhances the overall efficiency and resource utilization of the network. 

Diversity can be used in other ways. For instance, an SDN controller might choose to complete \textbf{both} e2e entanglements to improve the fidelity using distillation. Such an approach might improve the quality of entanglements, but both entanglements have to be created proximate in time so that one does not decohere before the other is created. The additional shared entanglement between end nodes can either be used to distill the entanglement, in which case we can afford a longer cut-off time $T$, or can be used to teleport two quantum states. In the results, we denote this as the ``\textit{all-completion}" policy where we wait for \textbf{both} the paths to have established e2e entanglement. 

Note here that for entanglement diversity, we would need one more unit of memory at the end nodes to store qubits in such a case that the link is established by both paths.

%\vspace*{3mm}
\subsection{Simulation}

We analyze the e2e entanglement generation process using simulations. For the simulations, we utilize the \texttt{networkX} package to model the quantum network. In accordance with our assumption that classical links are already present between all nodes, we do not explicitly add these links to the network. Instead, an edge between a pair of nodes strictly represents the presence of a shared entanglement. For the 4-hop path, we initialize the network by adding edges between nodes 1-2 and 3-5, as illustrated in Fig.~\ref{fig:fourhops}. This setup reflects the presence of prior entanglements. In contrast, the 2-hop path starts with no prior entanglements and must establish all links from scratch. In each time step, we perform the following operations:
\begin{itemize}
    \item \textbf{Entanglement Generation:} We attempt to generate entanglements for all pairs of nodes that do not already share an entanglement. An edge is successfully created if a randomly chosen number is less than $p_e$.
    \item \textbf{Entanglement Swapping:} Swaps are attempted between entanglements created in the previous time step, following the sequence shown in Fig.~\ref{fig:swaps}. A swap is successful if a randomly chosen number is less than $p_s$. If the swap fails, both entanglements involved in the swap are discarded.
    \item \textbf{Decoherence and Edge Discarding:} The fidelity of all existing edges is decreased by the amount specified by Eq.~\ref{F(t} to account for decoherence over one time step. If the fidelity of an edge falls below the threshold value $\mathcal{F}_{th}$, the edge is discarded. If the age of an edge exceeds the cutoff time $T$, the edge is discarded.
\end{itemize}
At the end of each simulation run, we record the total number of time steps required to establish e2e entanglement and the fidelity of the link between the end nodes for both the 2-hop and 4-hop scenarios. To ensure robust results, the simulation is repeated 10,000 times. Our distribution averages and quantiles are converged to within 2\% of the mean and quantile values.

\begin{figure*}[ht]
    \centering
    {\includegraphics[width=0.24\textwidth]{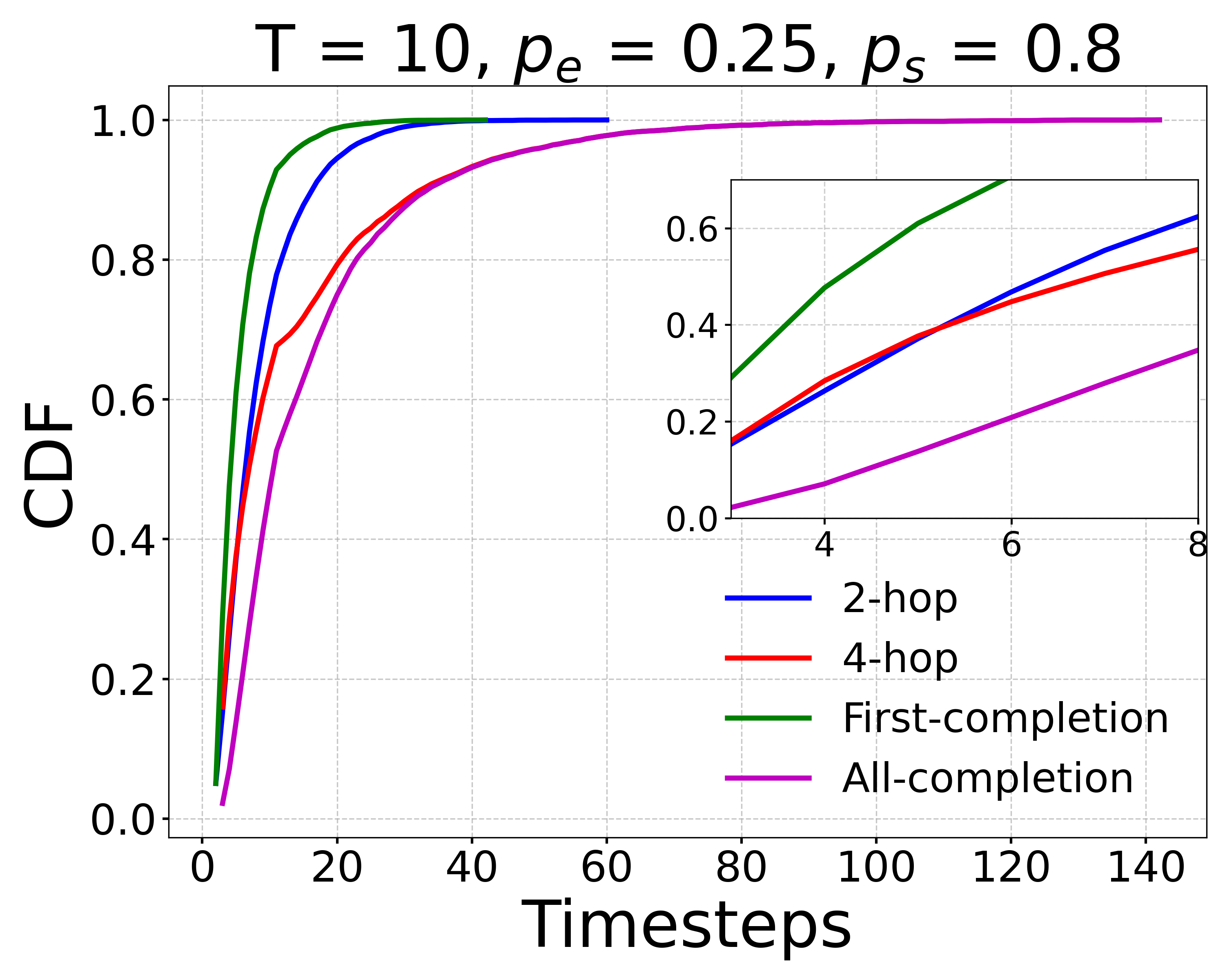}} 
    {\includegraphics[width=0.24\textwidth]{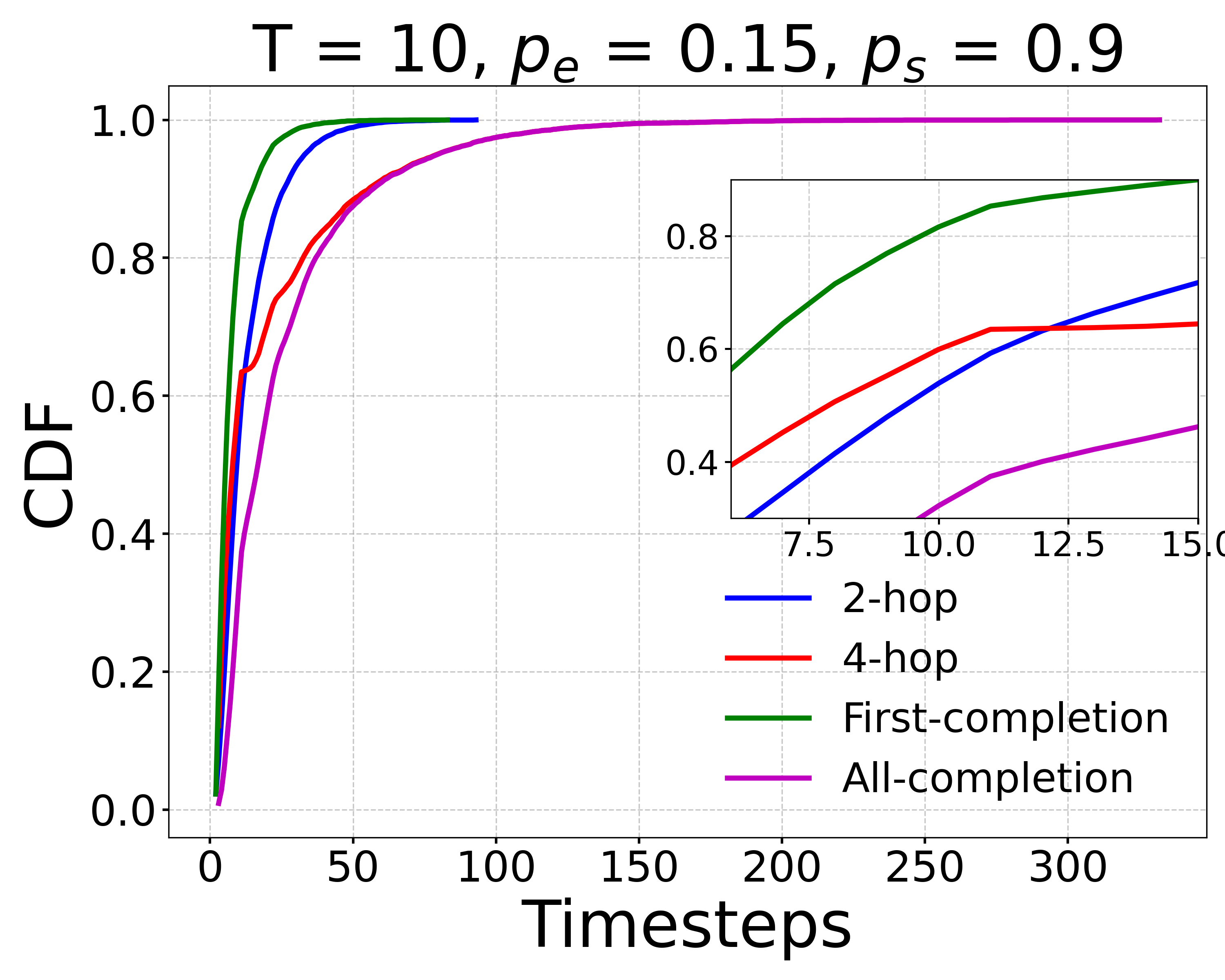}} 
    {\includegraphics[width=0.24\textwidth]{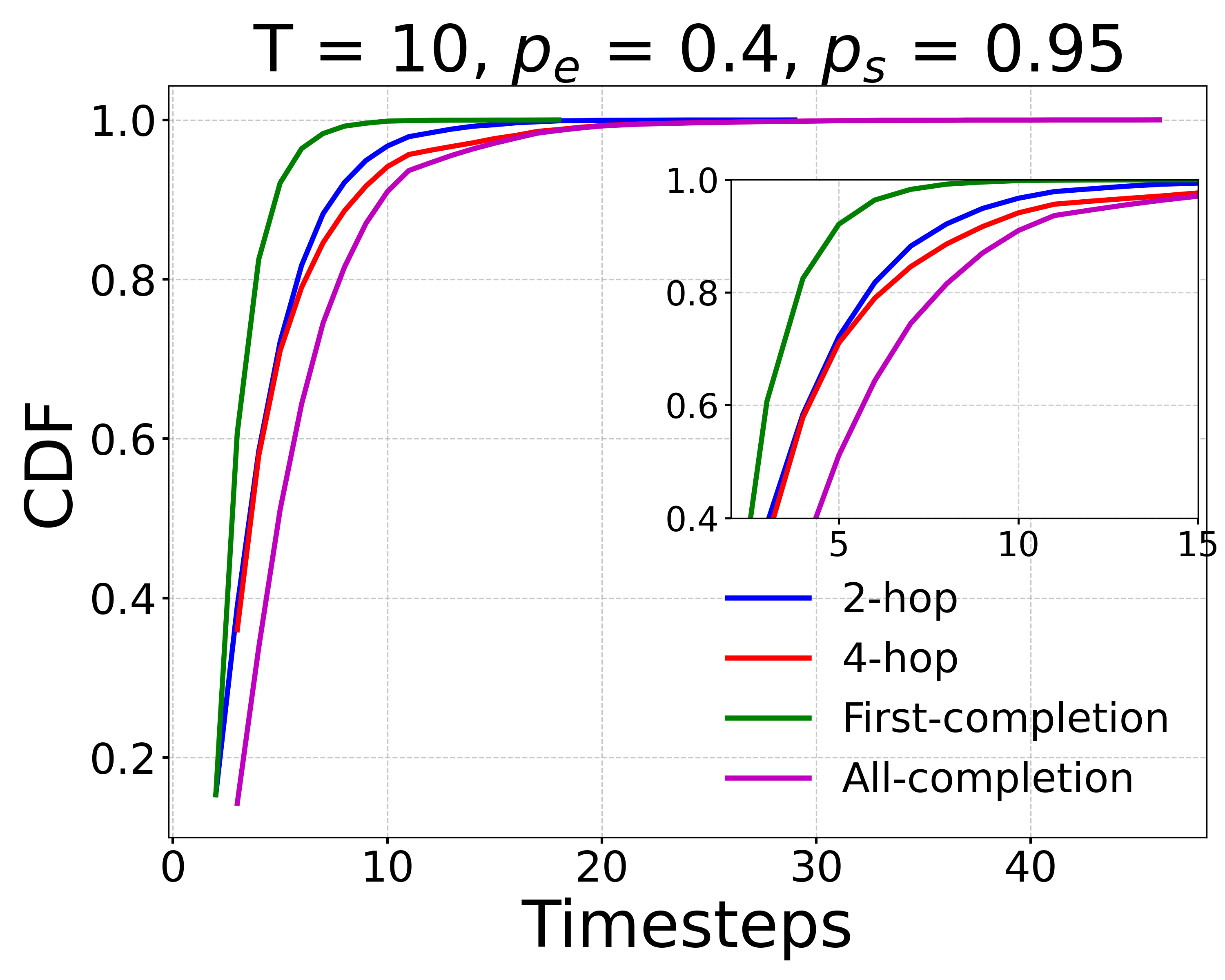}} 
    {\includegraphics[width=0.24\textwidth]{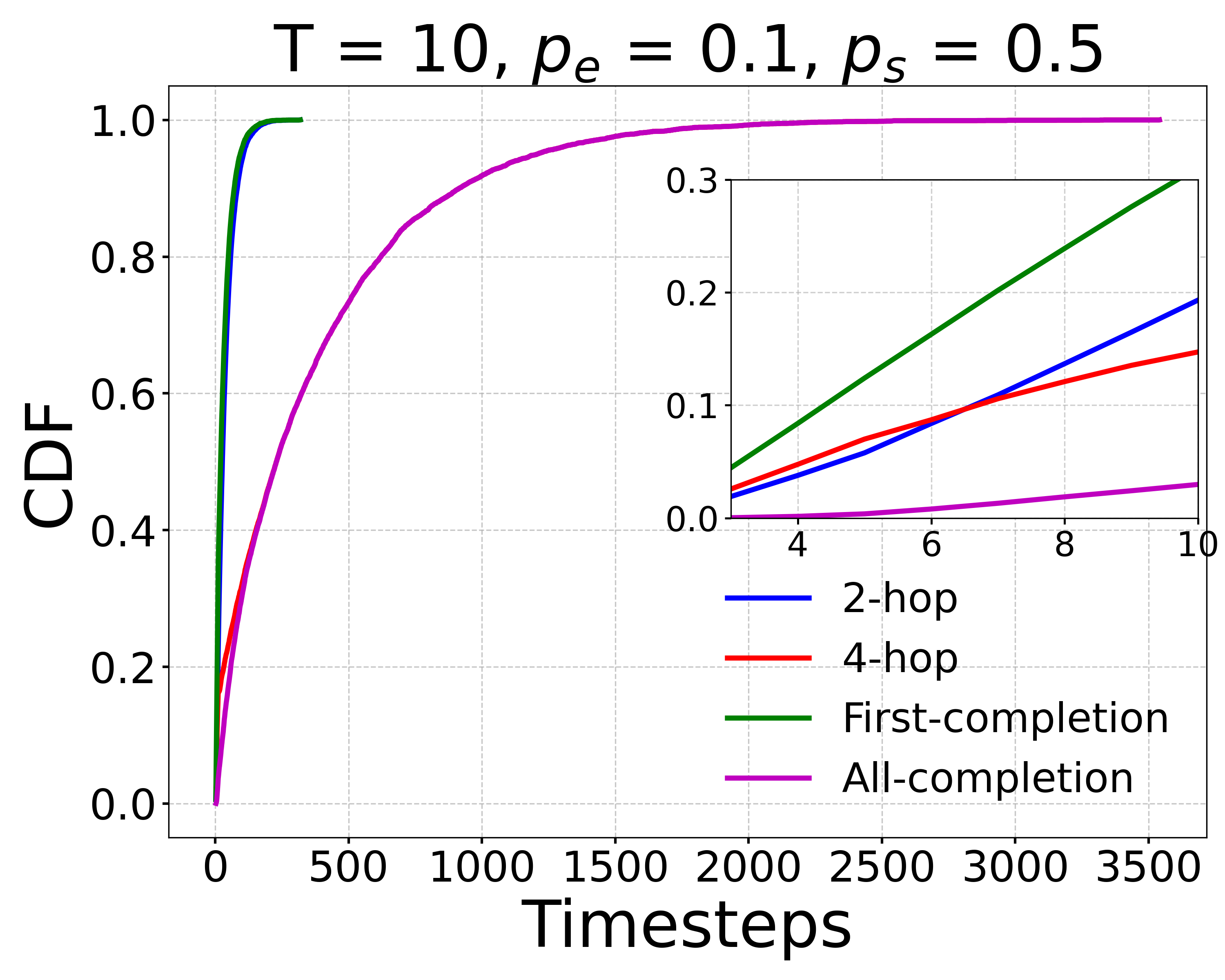}} 
    \caption{CDF of the number of time steps required for e2e entanglement generation for various values of $p_e$ and $p_s$ with $T$ = 10}
    \label{T=10 figures}
\end{figure*}

\section{Results}
%This metric is particularly critical as quantum entanglement (EPR pairs) serves as the foundational resource for quantum networking and the quantum internet but is highly susceptible to rapid degradation. Due to the no-cloning theorem, once entanglement fidelity degrades beyond usability, it is irretrievably lost. Hence, it is imperative to efficiently distribute and establish entanglement at the target nodes before decoherence renders it unusable.

In this section, we present a comprehensive analysis of the time required to establish e2e entanglement in quantum networks under varying system parameters. To evaluate the e2e entanglement generation performance, we compare a 2-hop and a 4-hop path under various values of the probability of successful entanglement generation, $p_e$, probability of successful swapping $p_s$, and cut-off time, $T$. The metric of interest is the number $N$ of discrete time steps required to achieve e2e entanglement. For each combination of $p_e$, $p_s$ and $T$, we compute the cumulative distribution function (CDF) of the distribution of the number of time steps needed to establish e2e entanglement. This is in contrast to the more commonly used average time to create e2e entanglement (e.g., \cite{caleffi2017optimal}) which is a crude characterization of the time it would take to create e2e entanglement. For example, a CDF value of $0.8$ at $N = 20$ indicates an 80\% chance that e2e entanglement is established within 20 time steps. %We analyze the performance of both 2-hop and 4-hop configurations across a range of parameter values, supported by contour plots that illustrate the CDF for fixed time steps as a function of $p_e$ and $p_s$. The CDF plots are added in the Appendix~\ref{appendix:contour_plots}. 

\subsection{Cutoff time of $T = 5$}

Based on Eq. \ref{F_th}, a cutoff time $T = 5$ corresponds to a high fidelity of 0.96. However, this stringent cutoff time of $T = 5$ significantly limits the performance of the 4-hop path. It is often too short to allow the 4-hop path to effectively utilize prior entanglements. In most cases, these prior entanglements decohere before they can be leveraged, forcing the 4-hop path to essentially start from scratch. Although there are conditions where the 4-hop path outperforms the 2-hop path, the advantages are marginal due to the challenge of utilizing prior entanglements before they decohere. Since the 4-hop path involves more intermediate steps without partial entanglements, it experiences greater delays once they decohere and is generally less efficient than the 2-hop path under these constraints.

We identify key thresholds where the 4-hop path provides an advantage. First, we find that $p_e = 0.25$ and $p_s = 0.8$ represent the upper and lower limits, respectively, for which the 4-hop path performs better than the 2-hop path as evidenced in Fig. \ref{T=5 figures}(a). We call this the ``\textit{4-hop favorable regime}". Specifically, for $p_e < 0.25$ and $p_s > 0.8$, the 4-hop path achieves e2e entanglement faster than the 2-hop path for the same CDF value within the first 6 time steps. For example, for $p_e = 0.15$ and $p_s = 0.9$ depicted in Fig. \ref{T=5 figures}(b), the 4-hop path maintains the lead. The primary result of interest here is that $p_e = 0.25$ and $p_s = 0.8$ are critical thresholds beyond which the 4-hop path no longer provides meaningful benefits.

Second, when $p_e$ increases to 0.4 with $p_s = 0.95$, the 2-hop path consistently overlaps with the 4-hop path, as shown in Fig. \ref{T=5 figures}(c). Note that $p_e = 0.4$ is outside the \textit{4-hop favorable regime} while $p_s = 0.95 > 0.8$. Initially, the 4-hop path keeps pace but degrades after six time steps due to decoherence, which negates any advantage from prior entanglements. The key limiting factor for the 2-hop path is the requirement to generate two entanglements within a constrained timeframe. In contrast, the 4-hop path only needs to generate one entanglement to utilize prior entanglements. However, for the 2-hop path to succeed, both entanglements must be created sufficiently close in time to prevent decoherence and ensure that the fidelity after the swap operation remains above the threshold $\mathcal{F}_{th}$. At $p_e=0.4$, the probability of generating both entanglements is high enough for the 2-hop path to reliably create these two entanglements within the required timeframe. On the other hand, the 4-hop path must generate its single entanglement quickly enough to leverage prior entanglements before they decohere. For $p_e>0.4$, the 2-hop path outperforms the 4-hop path because it can generate two entanglements fast enough to meet the fidelity and timing requirements, while the 4-hop path struggles to utilize its prior entanglements effectively with a short cutoff time $T$. Thus, $p_e > 0.4$ serves as a ``\textit{critical value of $p_e$}" for which the 2-hop path is always preferable. 

Third, for $p_e = 0.1$ and $p_s = 0.5$, the two cases nearly overlap in the CDF plot as seen in Fig. \ref{T=5 figures}(d). In this case, it is $p_s$ that is outside the \textit{4-hop favorable regime}. Even with a low $p_e = 0.1 < 0.25$, reducing $p_s$ below 0.5 is detrimental for the 4-hop path. In other words, for $p_s \leq 0.5$ as the ``\textit{critical value of $p_s$}", the 4-hop path cannot outperform the 2-hop path, regardless of how low $p_e$ is. This is primarily because low swap success probability disproportionately impacts the 4-hop path due to its greater number of intermediate steps. If a swap operation fails at any intermediate step, the corresponding entanglements are lost and need to be regenerated. However, if the final swap fails, the 4-hop path is at a significant disadvantage, as it must start from scratch. Consequently, the 4-hop path is more vulnerable to swap failures, and its performance degrades more severely compared to the 2-hop path when $p_s$ is low. This explains why the 4-hop path cannot surpass the 2-hop path for $p_s \leq 0.5$, even under favorable entanglement generation probabilities.

Finally, we observe that beyond 6 time steps, the \textit{all-completion} policy (both e2e entanglements are successful) consistently aligns with the 4-hop case, while the \textit{first-completion} policy (earliest e2e entanglment to succeed) overlaps with the 2-hop case. This demonstrates that once the 4-hop path loses the advantage of prior entanglements, it invariably generates e2e entanglement more slowly than the 2-hop path. Although higher values of $p_e$ and $p_s$ reduce the required time steps, the strict cutoff time of $T=5$ prevents the 4-hop path from leveraging prior entanglements effectively.

\begin{figure*}[ht]
    \centering
    {\includegraphics[width=0.24\textwidth]{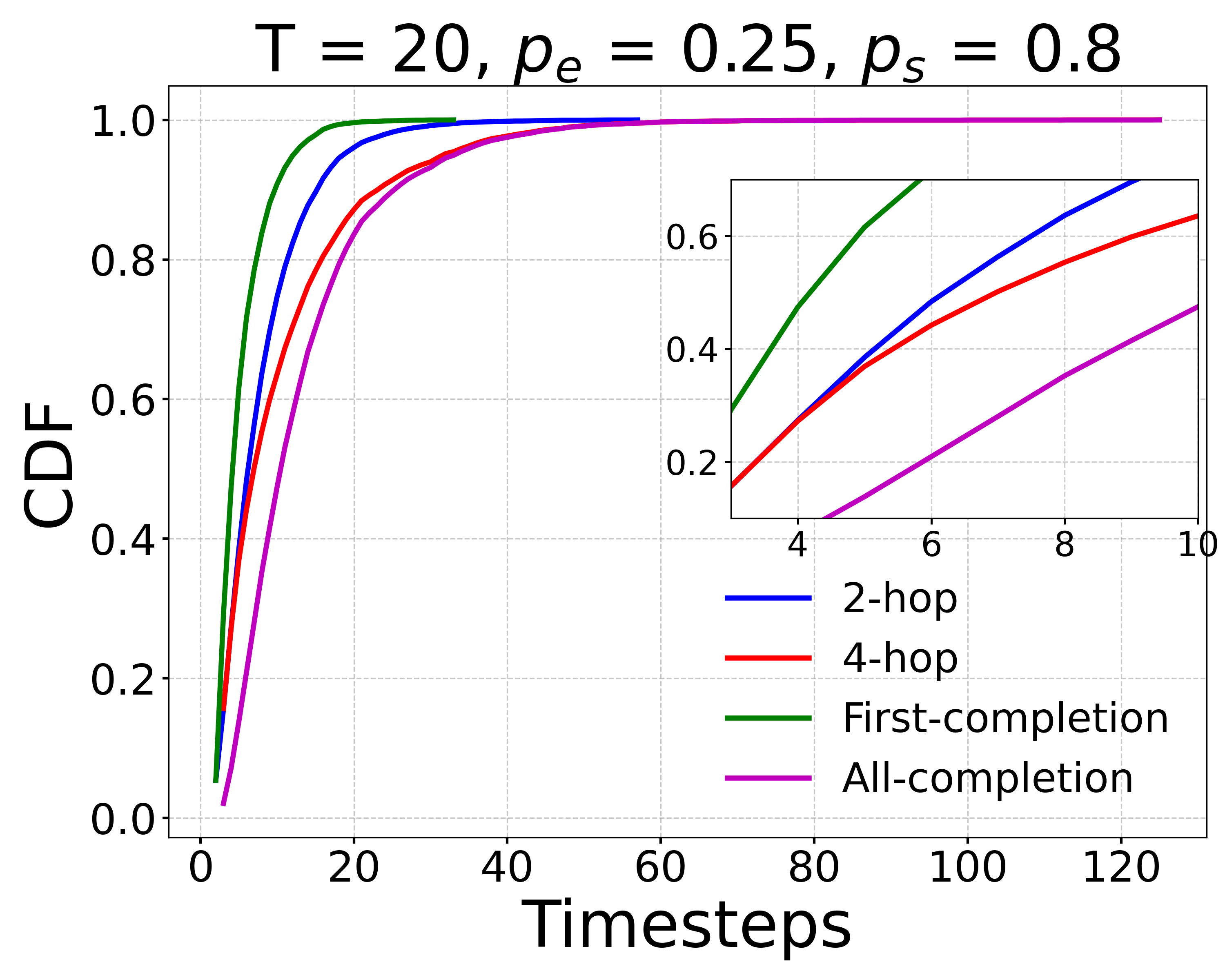}} 
    {\includegraphics[width=0.24\textwidth]{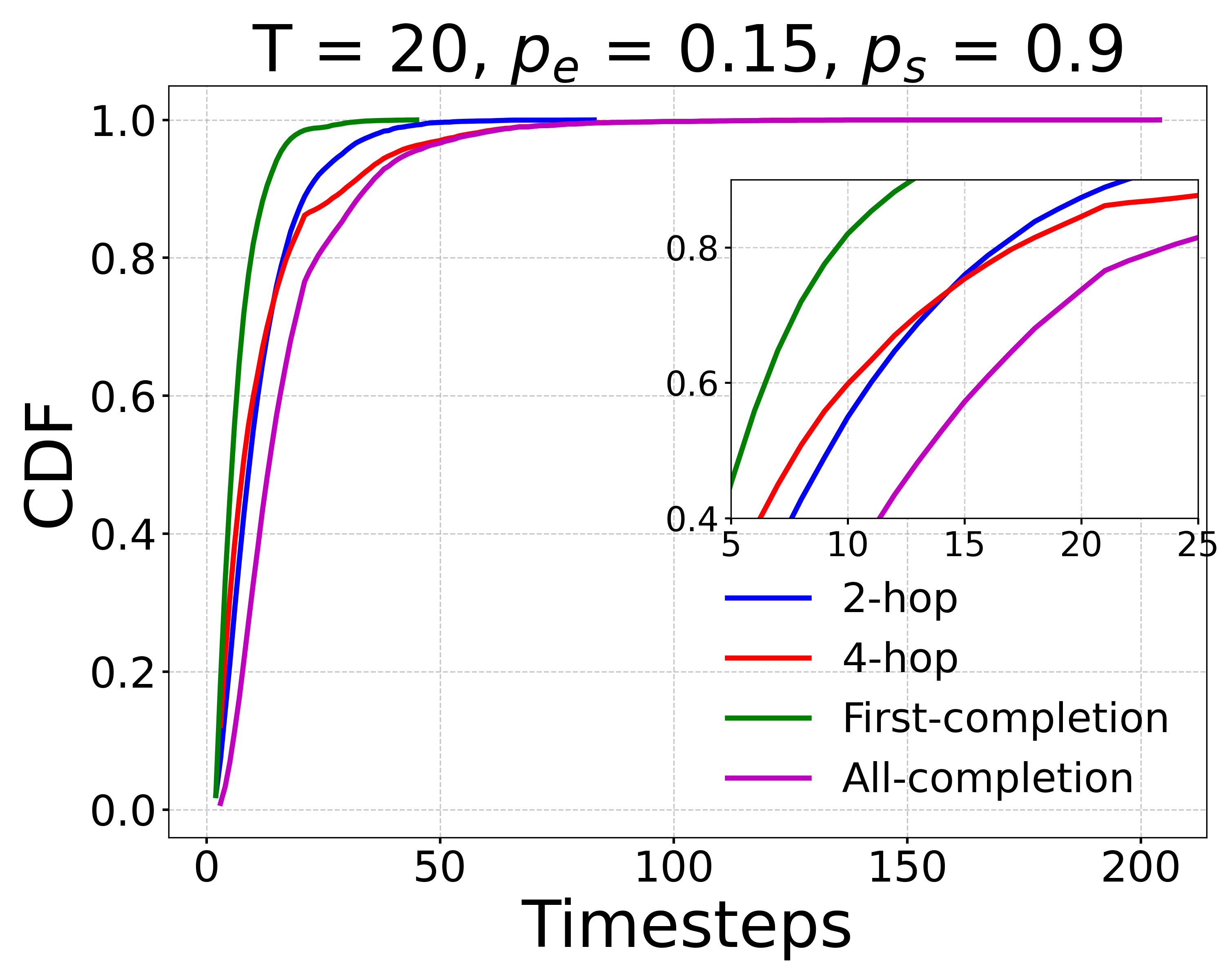}} 
    {\includegraphics[width=0.24\textwidth]{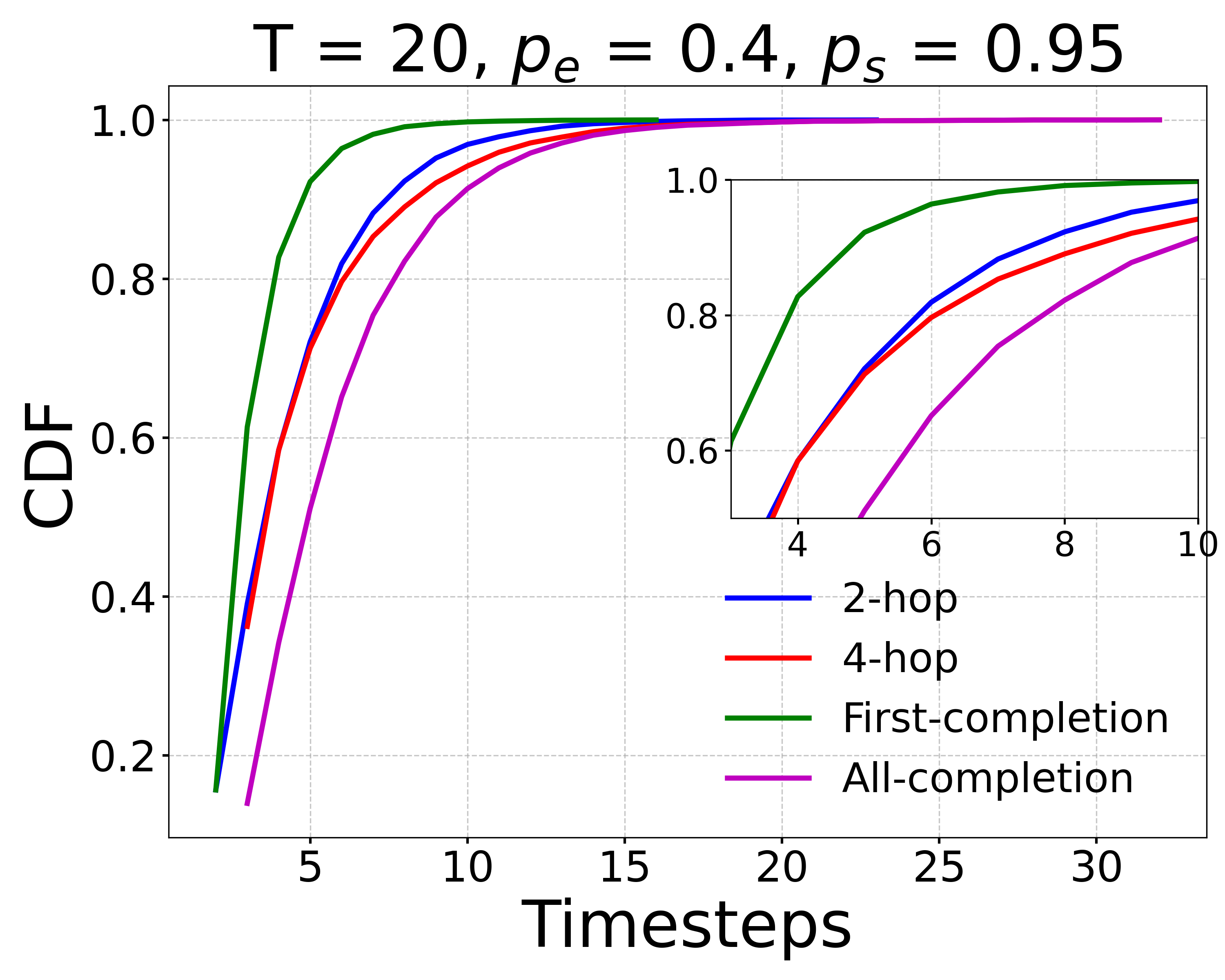}} 
    {\includegraphics[width=0.24\textwidth]{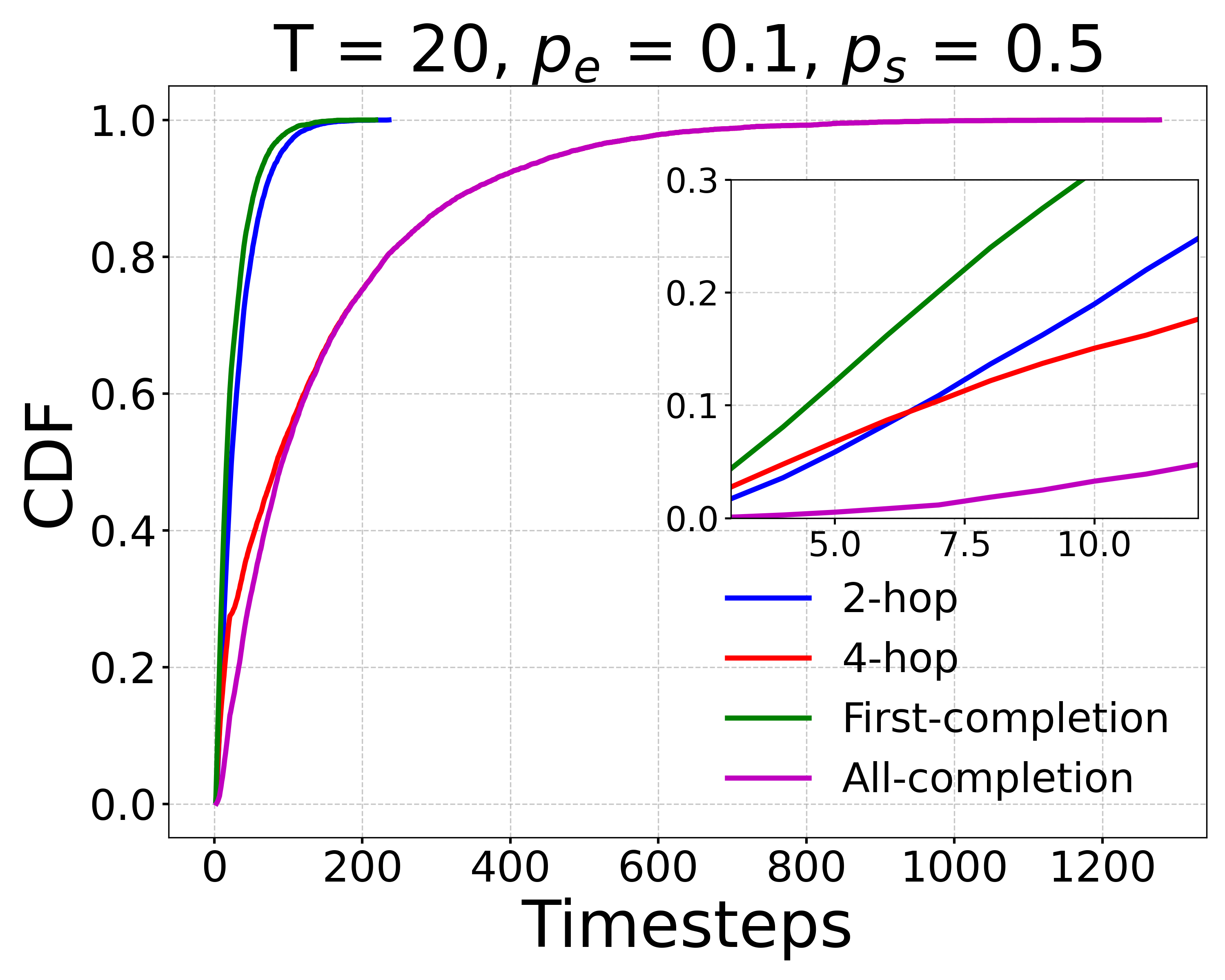}} 
    \caption{CDF of the number of time steps required for e2e entanglement generation for various values of $p_e$ and $p_s$ with $T$ = 20}
    \label{T=20 figures}
\end{figure*}

\subsection{Cutoff time of $T = 10$}

The critical values remain consistent with the case of $T = 5$. Specifically, the 4-hop path outperforms the 2-hop path in the \textit{4-hop favorable regime} with $p_e < 0.25$ and $p_s > 0.8$, as shown in Fig.~\ref{T=10 figures}(a). In contrast, the 2-hop path is always preferable for $p_e>0.4$ shown in Fig.~\ref{T=10 figures}(c), and for $p_s<0.5$, demonstrated in Fig.~\ref{T=10 figures}(d). These thresholds are independent of the cutoff time $T$ because $T$ only determines the duration after which links are discarded. While $T$ influences the number of time steps required to achieve e2e entanglement, it does not affect the intrinsic values of $p_e$ and $p_s$ which depend on the underlying qubit architecture.

As $T$ increases, the number of time steps required to achieve e2e entanglement decreases significantly, even though the critical value of $p_e$ and $p_s$ remain unchanged. For instance, at $p_e = 0.15$ and $p_s = 0.9$, the CDF value for the 4-hop path exceeds 0.6 within 10 time steps, compared to below 0.4 for $T = 5$. A longer cutoff time allows more opportunities to utilize pre-existing entanglements before they decohere, reducing the need for frequent probabilistic re-attempts and thereby improving overall efficiency. This effect is more pronounced for low $p_e$ and $p_s$, where successfully creating entanglements and performing swaps was already less likely. With $T = 5$, these scarce resources were often lost to decoherence, requiring more trials and thus time to recreate them. Extending $T$ to 10 mitigates this issue, significantly reducing the time steps required for both the 2-hop and 4-hop paths. However, this comes at the cost of reduced fidelity, as described by Eq.~\ref{F_th}, where $T = 10$ corresponds to a fidelity of 0.92. This highlights a trade-off between reducing the entanglement rate and maintaining high fidelity, which must be considered in practical implementations.

We also observe distinct performance trends similar to the case with $T=5$. For low $p_e$ and $p_s$, the \textit{all-completion} policy aligns with the 4-hop path, while the \textit{first-completion} policy corresponds to the 2-hop path, aside from the initial advantage of the 4-hop path. Again, the 4-hop path is more susceptible to decoherence and swap failures due to its greater number of intermediate steps. The advantages of prior entanglements are quickly lost as entanglements decohere even with their longer lifespan before final swapping, forcing the 4-hop path to restart frequently. In contrast, the 2-hop path, with fewer steps, is less affected by decoherence and can recover more efficiently from failures. For intermediate and high values of $p_e$ and $p_s$, both paths initially fall between the \textit{first}- and \textit{all-completion} policies' plots. However, beyond a certain point, the 4-hop path converges with the \textit{all-completion} policy's performance, while the 2-hop path aligns with the \textit{first-completion} policy. %This behavior arises because the 4-hop path’s additional steps introduce more opportunities for failures and decoherence over time, while the 2-hop path’s simpler structure allows it to maintain efficiency and reliability. 
Thus, outside the favorable regime ($p_e < 0.25$ and $p_s > 0.8$), the 2-hop path is strictly superior due to its robustness against decoherence and swap failures, as well as its ability to efficiently utilize entanglement resources within the cutoff time.

\subsection{Cutoff time of $T = 20$}

We observe similar threshold behavior for the $T = 20$ case as in the $T = 5$ and $T = 10$ scenarios. The 4-hop path remains the better choice within the \textit{4-hop favorable regime} with $p_e \leq 0.25$ and $p_s \geq 0.8$, as shown in Fig.~\ref{T=20 figures}(a). For $p_e \geq 0.4$ and $p_s = 0.95$, the performance of the 2-hop and 4-hop paths nearly overlaps in the CDF plot, as seen in Fig.~\ref{T=20 figures}(c) and the 4-hop path cannot outperform the 2-hop path, regardless of the value of $p_s$ suggesting that $p_e = 0.4$ is still the \textit{critical value of $p_e$}. Similarly, $p_s \leq 0.5$ remains the \textit{critical value of $p_s$} for which the 4-hop path cannot outperform the 2-hop path, regardless of the value of $p_e$, as shown in Fig.~\ref{T=20 figures}(d). 

A notable observation is the presence of a crossover point, where the 4-hop path initially performs better but is eventually overtaken by the 2-hop path. This crossover point occurs at an earlier time step than the cutoff time $T$. The cutoff time $T$ determines the maximum duration a link can remain unused before being discarded, but fidelity constraints impose an additional limit. A link is discarded if its fidelity decays below the threshold $\mathcal{F}_{th}$, which, in the absence of usage, would occur exactly after $T$ time steps. However, due to noisy entanglement swaps and fidelity decay in quantum memories, the threshold fidelity $\mathcal{F}_{th}$ is often reached before $T$, which explains why the crossover point occurs earlier than $T$. 

Furthermore, the crossover point occurs at fewer time steps for parameter values beyond the \textit{4-hop favorable regime}. %($p_e \geq 0.4$ or $p_s \leq 0.8$). 
For example, in Fig.~\ref{T=20 figures}(b), the crossover point occurs at approximately 15 time steps for $T = 20$. However, for $p_e \geq 0.4$ or $p_s \leq 0.8$, the crossover point shifts to around 5 time steps. This is because the 4-hop path performs poorly in these regimes, losing its advantage sooner, and thus the 2-hop path overtakes it more quickly. 

Lastly, based on Eq.~\ref{F_th}, a cutoff time of $T = 20$ corresponds to a fidelity of 0.86. While increasing $T$ reduces the number of time steps required to achieve e2e entanglement, it also lowers the fidelity of the e2e entanglement. This reinforces the importance of balancing fidelity constraints with the rate of entanglement generation when designing quantum networks.

\begin{figure*}[ht]
    \centering
    {\includegraphics[width=0.24\textwidth]{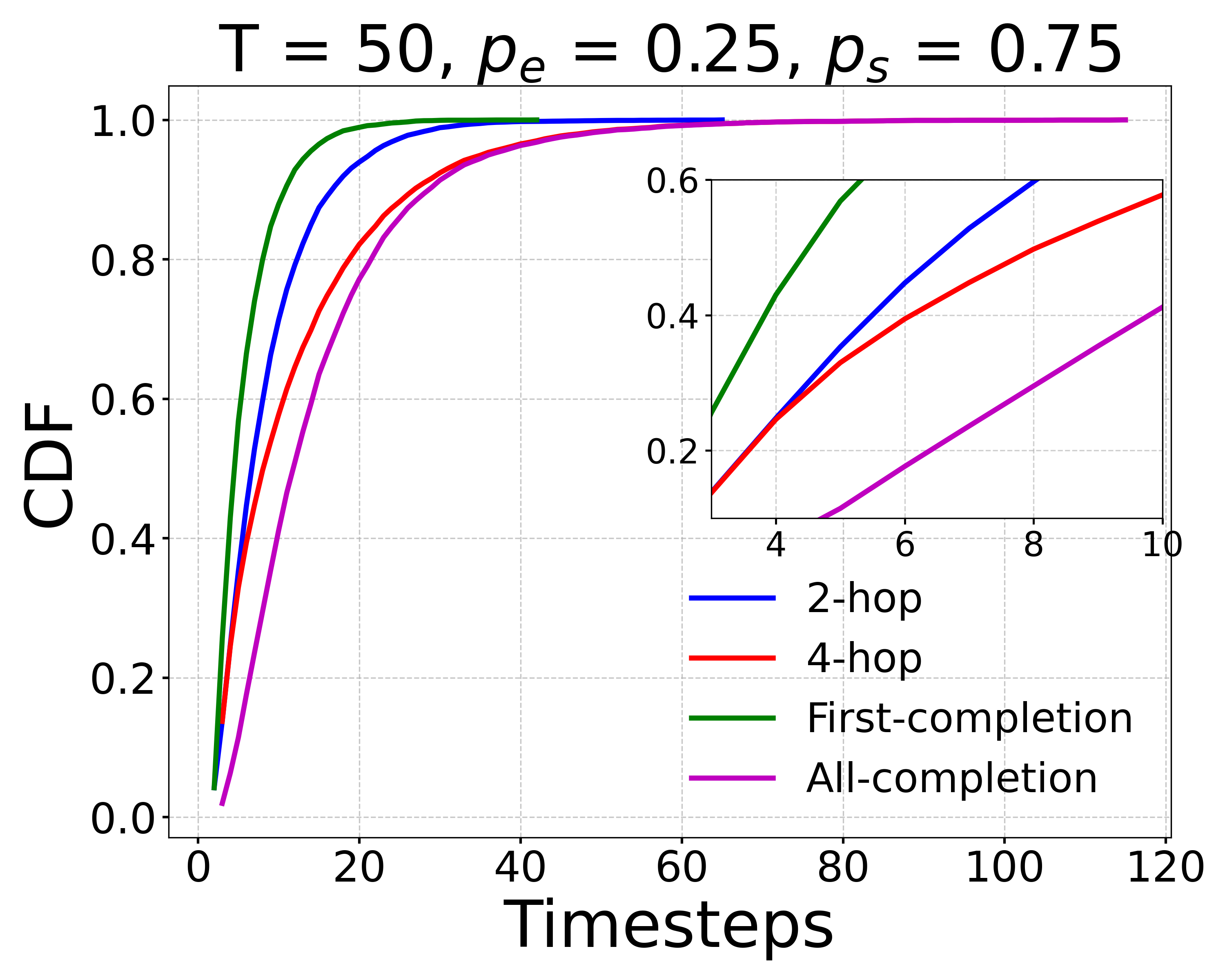}} 
    {\includegraphics[width=0.24\textwidth]{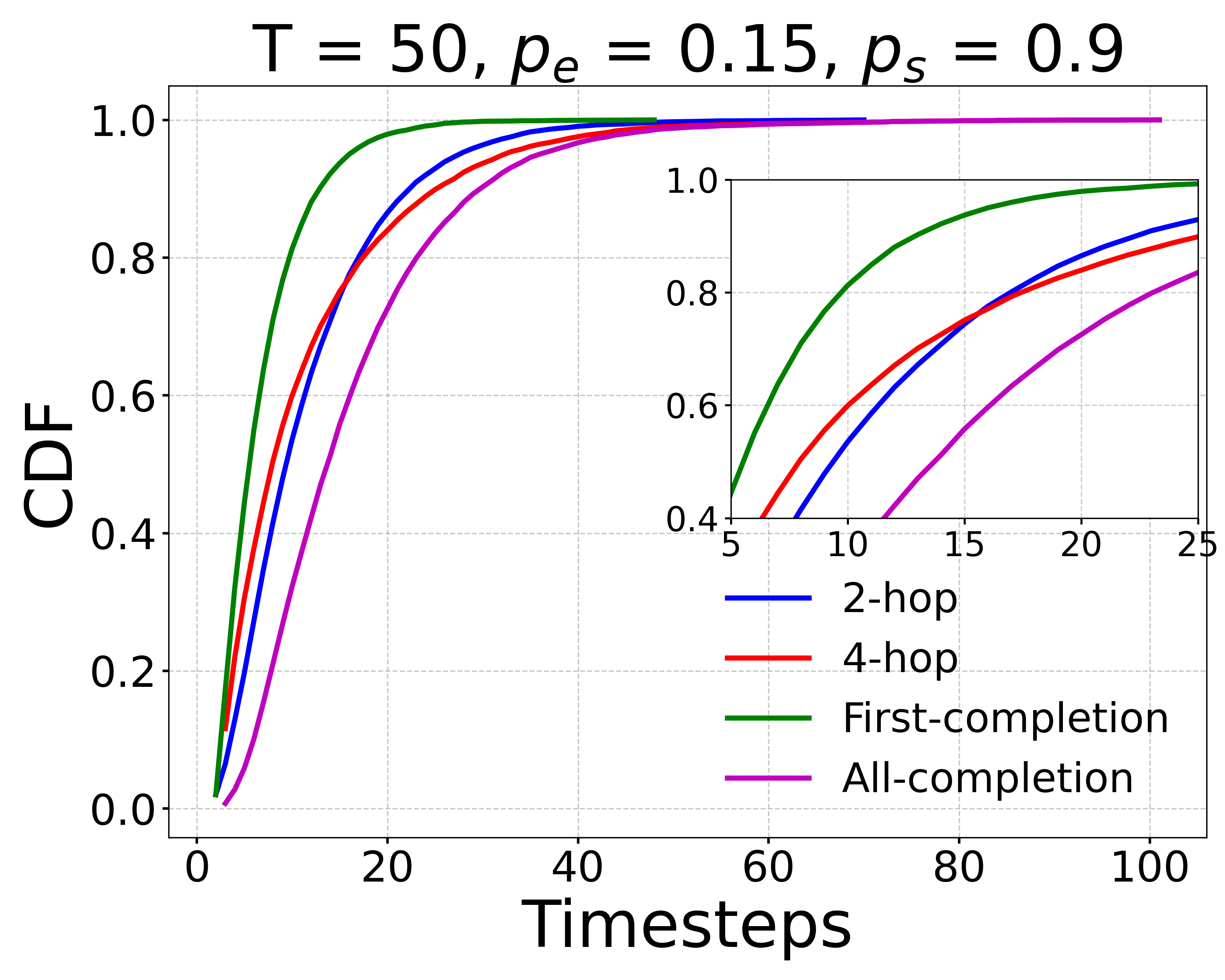}} 
    {\includegraphics[width=0.24\textwidth]{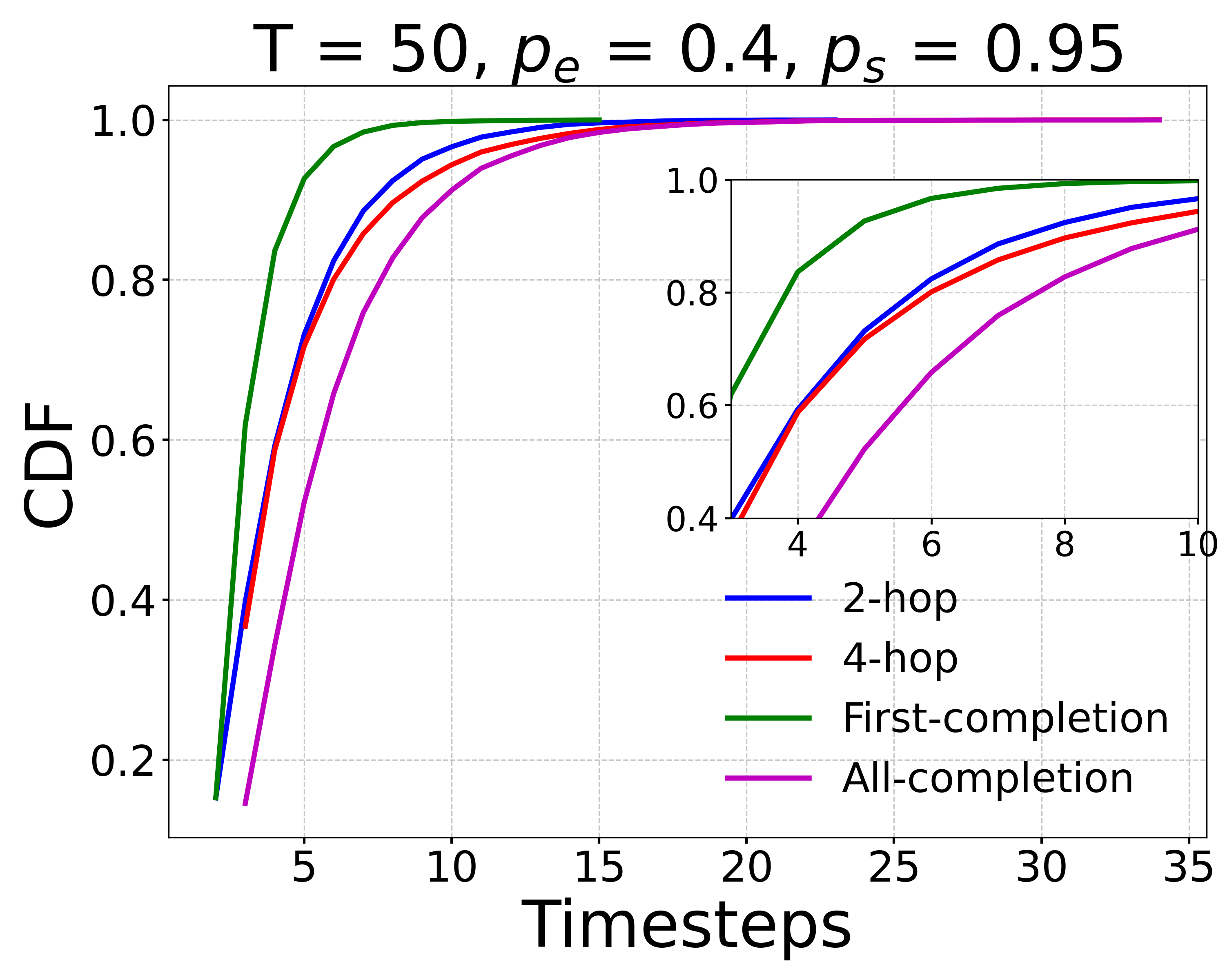}} 
    {\includegraphics[width=0.24\textwidth]{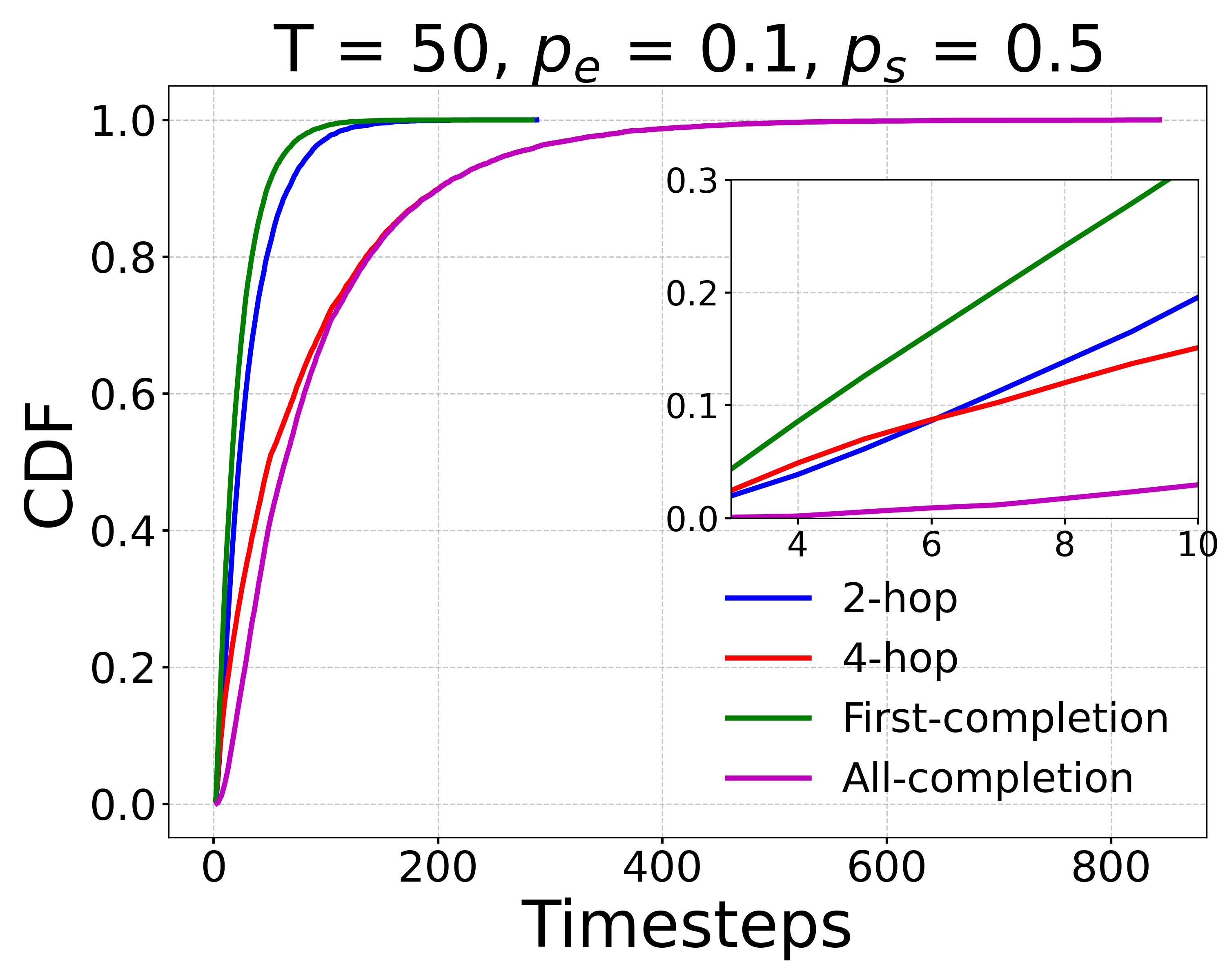}} 
    \caption{CDF of the number of time steps required for e2e entanglement generation for various values of $p_e$ and $p_s$ with $T$ = 50}
    \label{T=50 figures}
\end{figure*}

\subsection{Cutoff time of $T = 50$}

We observe similar trends for the $T = 50$ case, with slight adjustments to the thresholds. The 4-hop path remains the better choice for $p_e \leq 0.25$ and $p_s \geq 0.75$, a slight relaxation from the previous \textit{4-hop favorable regime} of $p_s \geq 0.8$, as shown in Fig.~\ref{T=50 figures}(a). However, even with this extended cutoff time, there is no noticeable improvement in the CDF value for e2e entanglement within 10 time steps for the 4-hop path, as demonstrated for $p_e = 0.15$ and $p_s = 0.9$ in Fig.~\ref{T=50 figures}(b). This suggests that the benefits of increasing the cutoff time diminish as $T$ grows larger, likely due to the saturation of gains from pre-existing entanglements and the persistent impact of decoherence on stored entanglements. Beyond a certain point, increasing $T$ does not significantly improve performance, as the pronounced effect of swap failures followed by new probabilistic entanglement generation continues to limit the 4-hop path's ability to generate e2e entanglements.

For $p_e = 0.4$ and $p_s = 0.95$, the performance of the 2-hop and 4-hop paths continues to overlap in the CDF plot, as seen in Fig.~\ref{T=50 figures}(c). Further increasing the cutoff time does not provide a meaningful advantage for this threshold, aside from reducing the total number of time steps and marginally increasing the CDF values. This reaffirms that for $p_e \geq 0.4$ as the \textit{critical value of $p_e$}, the 4-hop path cannot outperform the 2-hop path, regardless of the value of $p_s$. Similarly, for $p_s \leq 0.5$ as the \textit{critical value of $p_s$}, the 4-hop path remains inferior to the 2-hop path, irrespective of $p_e$, as shown in Fig.~\ref{T=50 figures}(d). 

We also observe that the crossover point occurs at significantly fewer time steps than the cutoff time $T$. For example, in the $T = 50$ case, the crossover point occurs at approximately 16 time steps for $p_e = 0.15$ and $p_s = 0.9$, as shown in Fig.~\ref{T=50 figures}(b). This behavior is consistent with earlier observations and can be attributed to the fidelity decay due to noisy swaps as explained in the previous subsection. Due to noisy entanglement swaps and fidelity decay in quantum memories, the threshold fidelity $\mathcal{F}_{th}$ is often reached well before $T$, causing the crossover point to occur earlier than the cutoff time. This effect is more pronounced for parameter values beyond the favorable regime, where the crossover point shifts to as few as 5 time steps. In these regimes, the 4-hop path loses its initial advantage rather quickly, and the 2-hop path performs comparably better due to its simpler structure and steps to e2e entanglement.

Furthermore, the trends for the \textit{first}- and \textit{all}-completion policies remain consistent across $T = 5,$ $10,$ and $20$. For low values of $p_e$ and $p_s$, the \textit{all-completion} policy's performance aligns with the 4-hop path, while the \textit{first-completion} policy's performance corresponds to the 2-hop path, aside from the initial advantage or crossover point of the 4-hop path. For intermediate and high values of $p_e$ and $p_s$, both paths initially fall between the \textit{first}- and \textit{all}-completion policies' plots. This is because, in these regimes, the two paths exhibit comparable performance, making it difficult to determine which one achieves e2e entanglement first. Since the results are probabilistic, both paths achieve e2e entanglement in similar timeframes, leading to an initial period where neither is strictly superior.
However, beyond a certain point, the 4-hop path converges with the \textit{all-completion} policy's performance, while the 2-hop path aligns with the \textit{first-completion} policy. This transition occurs because the 2-hop path has a narrower range of e2e entanglement generation times, allowing its CDF to rapidly approach 1. In contrast, the 4-hop path, with its additional steps when the prior entanglements are lost, exhibits a wider range of generation times with a significant portion of attempts taking considerably longer. As a result, by the time the 2-hop path's CDF approaches 1, the 4-hop path is still generating e2e entanglement in longer times, causing it to merge with the \textit{all-completion} policy's performance.

Finally, the trade-off between cutoff time and fidelity continues to be a crucial factor. Based on Eq.~\ref{F_th}, a cutoff time of $T=50$ corresponds to a further reduction in fidelity compared to previous cases. While increasing $T$ helps reduce the number of time steps required for entanglement generation, it comes at the cost of a lower final fidelity.

\section{Analysis}
In our simulations, we have chosen a coherence time of $100$ ms for the quantum memory, which is a modest estimate chosen to remain general and inclusive of all possible qubit architectures. This choice ensures that our results are applicable to a wide range of quantum memory platforms, including those with shorter coherence times. Based on the numerics given in \cite{caleffi2017optimal}, the time for one time step $\Delta t$ is $106\mu s$ which is consistent with the operational timescales of various qubit architectures. However, it is important to note that long coherence times have been experimentally demonstrated for various quantum memory systems, such as longer than hundreds of seconds for trapped ions \cite{langer2005long, haffner2005robust}, and over a second for optical memories \cite{ranvcic2018coherence} and atom-based quantum memories \cite{harber2002effect, treutlein2004coherence}. For such high-performance memories, the fidelity decay during qubit storage is significantly reduced, leading to fewer discarded links and a corresponding decrease in the number of time steps required to achieve e2e entanglement.

\begin{figure}[hth!]
    \centering
    \includegraphics[width=0.85\columnwidth]{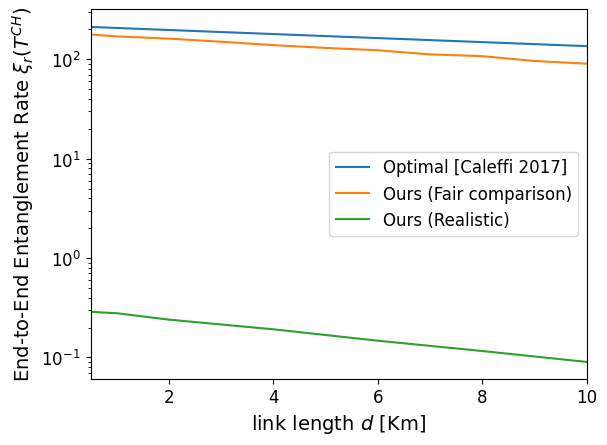}
    \caption{Comparison with the average e2e entanglement generation rate presented in \cite{caleffi2017optimal}}
    \label{caleffi comparison}
\end{figure}

We compare our work with a 4-hop path presented in \cite{caleffi2017optimal}, using the average end-to-end entanglement rate expression derived in their study. As shown in Fig. \ref{caleffi comparison}, our results are plotted under two different settings: \textbf{ours (realistic)}, which incorporates realistic system parameters and all the physical constraints mentioned throughout this paper, and \textbf{ours (fair comparison)}, which uses similar parameter values and assumptions as in \cite{caleffi2017optimal} for a direct comparison. Specifically, we set $p_s=1$ to reflect their assumption that links are not discarded when a swap fails, which effectively models a perfect swapping operation. We also lower the fidelity threshold to $ F_{\text{th}} = 0.8 $ to account for their omission of noise in entanglement generation and swapping and lack of fidelity decay in quantum memories. Under these conditions, our entanglement rate closely matches that of the optimal routing scheme reported in their paper. However, when realistic noise and loss are included, our end-to-end entanglement rate is nearly three orders of magnitude lower. This demonstrates the importance of incorporating physical effects when analyzing and comparing routing strategies in quantum networks.

In our work, we assumed the age of the oldest link when performing a swap. By calculating the exact age of the link established through entanglement using the fidelity decay equation (Eq.~\ref{F(t}), we can further improve our e2e entanglement rate or reduce the number of time steps required to achieve e2e entanglement. Additionally, we have not optimized the timing of entanglement swaps, as this was not the primary focus of our study. However, such optimization could be explored in future work to improve the e2e entanglement generation rate.

It is important to note that our work cannot be directly compared with most existing studies, as none of them comprehensively account for all the physical constraints incorporated in our model. For instance, \cite{li2022fidelity, li2021effective} ensure that a threshold fidelity is maintained for optimal routing but do not account for fidelity decay during storage in quantum memories or probabilistic $p_e$ and $p_s$. Similarly, \cite{chakraborty2019distributed} considers noisy memories but assumes deterministic swaps, while \cite{nguyen2022maximizing} assumes near-deterministic entanglement generation. Furthermore, \cite{victora2020purification} considers deterministic swaps and near-deterministic entanglement generation. In contrast, our framework explicitly incorporates fidelity decay, probabilistic entanglement generation, and swapping, providing a more realistic and comprehensive model.

\section{Discussion}
\label{sec:discussion}
The values for $p_e$ and $p_s$ are critical parameters in the design and performance of quantum networks. We identify \textit{critical values} of $p_e \leq 0.4$ and $p_s \geq 0.8$ as key regimes where the existence of prior entanglements becomes a crucial consideration in selecting the optimal path for e2e entanglement generation. Since the 4-hop path requires more swaps (2 swaps and 1 entanglement generation) compared to the 2-hop path (1 swap and 2 entanglement generations), higher $p_s$ is particularly advantageous for the 4-hop case. These thresholds are particularly significant for photonic qubits, as $p_e$ values are typically small due to distance-dependent attenuation in optical fibers. Furthermore, recent advances have shown improvements in the BSM success probability to $0.75$ and pushing $p_s$ closer to unity under certain conditions \cite{grice2011arbitrarily, ewert20143}. 
In addition to photonic qubits, trapped ion qubits entangled through single-rail or dual-rail photonic encodings \cite{drahi2021entangled} exhibit similar relationships for $p_e$. Trapped ion systems also benefit from deterministic entanglement swapping \cite{riebe2008deterministic}, while $p_e$ remains small due to challenges in photon collection and detection efficiencies \cite{blatt2008entangled, hauke2015probing}. This places trapped ion qubits squarely within the regime of high $p_s$ and low $p_e$, making the consideration of prior entanglements essential for optimizing path selection in such networks.

%For small values of $T$ (e.g., $T$= 5 ) or outside the favorable regime, the worst-case performance almost overlaps with the 4-hop path, while the best-case performance aligns with the 2-hop path. This suggests that outside the favorable regime, the 2-hop path is strictly better. These results have significant implications for decision-making in quantum networks, as they provide clear guidelines for selecting the optimal path based on network parameters and application requirements. 

From a practical perspective, our findings suggest that network designers must carefully consider not only the trade-offs between time to e2e entanglement and fidelity but also other realistic factors when selecting routes for entanglement generation. For time-sensitive applications, the CDF value can provide a useful estimate of which path to select if the application requires e2e entanglement within a given number of time steps. However, for applications that require high entanglement rates, such as distributed quantum computing, larger values of $T$ can be selected, and the choice of path can be tailored to the specific network or qubit architecture.

Additionally, \textit{entanglement diversity} is a useful strategy to minimize the number of time steps required for e2e entanglement. However, this approach depends on the available memory at network nodes. While our current work considers isolated demand, in a real network with multiple demands and varying edge capacities, the choice of a single path, entanglement diversity, or waiting for both paths to establish e2e entanglement for distillation will become critical.

We aim to generalize our framework to consider a full network with multiple demands, varying edge capacities, and internodal distances. This will enable the development of an optimal policy for path selection that incorporates all the constraints discussed here, along with the state of the network and prior entanglements after each iteration. Such a framework will provide a more comprehensive tool for designing and optimizing quantum networks in realistic scenarios.

\section{Conclusion}
In this work, we have demonstrated that by explicitly considering prior entanglements in a quantum network, longer paths (e.g., 4-hop) can, in certain regimes, establish e2e entanglement faster than shorter paths (e.g., 2-hop). This counterintuitive result highlights the importance of leveraging prior entanglements and carefully selecting paths based on network conditions. Our framework incorporates all necessary physical constraints to fully characterize a quantum network, including decoherence in quantum memories, probabilistic entanglement generation and swapping, and the discarding of links when swaps fail. By doing so, we provide a realistic and comprehensive model for analyzing and optimizing quantum networks.

We identified \textit{critical values} for the entanglement generation probability ($p_e$) and the entanglement swapping success probability ($p_s$) where the 4-hop path outperforms the 2-hop path. Specifically, for $p_e \leq 0.25 $ and $p_e \geq 0.8$, the 4-hop path benefits from prior entanglements and higher swapping success rates, making it the preferred choice in these regimes.

We have also shown that \textit{entanglement diversity}, enabled by memory at network nodes, offers a powerful strategy to minimize the time to e2e entanglement while maintaining fidelity above a specified threshold. This approach provides a valuable trade-off between rate and fidelity, making it particularly useful for time-sensitive applications. 

Our findings have significant implications for the design and operation of quantum networks. They provide clear guidelines for selecting optimal paths based on network parameters, such as $p_e$, $p_s$, and the cutoff time $T$. Additionally, our results underscore the importance of incorporating realistic constraints, such as fidelity decay and probabilistic operations, into network models to ensure accurate performance predictions.

\section*{Acknowledgement}
Authors Anoosha Fayyaz and Kaushik P. Seshadreesan acknowledge partial support from National Science Foundation under Grant No. 2204985.

\bibliographystyle{ieeetr}
\bibliography{references}

\end{document}